\documentclass[journal]{IEEEtran}

\ifCLASSINFOpdf
   \usepackage[pdftex]{graphicx}
\else
\fi

\usepackage{graphicx}
\usepackage[cmex10]{amsmath}
\usepackage{cite}

\begin{document}

\title{A Theory of Nonlinear Signal-Noise Interactions in Wavelength Division Multiplexed Coherent Systems}

\author{Amirhossein~Ghazisaeidi %,~\IEEEmembership{Member,~IEEE,}
        
\thanks{A. Ghazisaeidi is with Nokia Bell Labs, 
Paris-Saclay, Route de Villejust, 91620, Nozay, France
e-mail: (amirhossein.ghazisaeidi@nokia-bell-labs.com).}}

\markboth{March ~2017}
{Shell \MakeLowercase{\textit{et al.}}: }

\maketitle

\begin{abstract}
a general theory of nonlinear signal-noise interactions for wavelength division multiplexed fiber-optic coherent transmission systems is presented. This theory is based on the regular perturbation treatment of the nonlinear Schr\"odinger equation, which governs the wave propagation in the optical fiber, and is exact up to the first order in the fiber nonlinear coefficient. It takes into account all cross-channel nonlinear four-wave mixing contributions to the total variance of nonlinear distortions, dependency on modulation format, erbium-doped fiber and and backward Raman amplification schemes, heterogeneous spans, and chromatic dispersion to all orders; moreover, it is computationally efficient, being 2-3 orders of magnitude faster than the available alternative treatments in the literature. This theory is used to estimate the impact of signal-noise interaction on uncompensated, as well as on nonlinearity-compensated systems with ideal multi-channel digital-backpropagation.  
\end{abstract}


\IEEEpeerreviewmaketitle

\section{Introduction}

\IEEEPARstart{T}{he} spectral efficiency of fiber-optic wavelength division multiplexed (WDM) transmission systems is fundamentally limited by intra- and inter-channel four-wave mixing (FWM) processes stemming from the optical fiber intensity-dependent nonlinear Kerr refractive index \cite{Agrawal book, Essiambre2010 : JLT2010}. Many of the recent high-capacity ultra long-haul ``hero experiments" exploited single-channel digital nonlinear compensation (NLC) to deal with fiber nonlinear impairments and to push the transmission limits imposed by such FWM terms \cite{Ghazisaeidi2014 : ECOC2014 PDP, Ghazisaeidi2015 : JLT2015, Ghazisaeidi2015 : ECOC2015, Ghazisaeidi2016 : JLT2016, Ghazisaeidi2016 : ECOC2016, Ghazisaeidi2017 : JLT2017, NEC2016 : OFC2016, SubCom2017 : OFC2017}. 

The propagation of the perfectly polarized electromagnetic field in the optical fiber is modeled by the nonlinear Schr\"odinger equation (NLS). In order to account for the random birefringence in fiber optics, it is common to use the Manakov approximation of the system of two coupled NLS equations governing the transverse components of the total electromagnetic field\cite{Agrawal book}.

Since 2010, much progress has been made in developing analytical models for polarization-multiplexed (PM) WDM coherent transmission systems. These models have been proved to be particularly successful in predicting the performance of dispersion un-managed (DU) systems, and have been validated numerically and experimentally many times by various groups. (for instance cf. \cite{Vacondio2012: OptXpress2012}). Most of these analytical models are based on the regular perturbation (RP) approximation of the solutions of NLS and/or Manakov equation, where only terms up to the first-order in fiber nonlinear coefficient are kept. 

The first attempts in developing a complete theory of nonlinear propagation in the modern coherent PM-WDM DU systems, by Chen, Poggiolini and Carena, resulted in the so-called Gaussian noise model (GNM)\cite{Chen2010 : OptXpress2010, Poggiolini2011 : PTL2011, Carena2012 : JLT2012, Poggiolini2014 : JLT2014}. An independent derivation was published by Johannisson and Karlsson \cite{Johannisson2013 : JLT2013}. In the GNM, the WDM signal is modeled as a Gaussian random process all along the link, \emph{i.e.}, from the injection point into fiber at the transmitter side up to the receiver side front-end. This Gaussian random process is represented as a grid of Dirac delta functions in the frequency domain at the channel input. The amplitude of each delta function is modulated by a Gaussian random variable. All FWM terms among these spectral lines are then computed at channel output, and finally the power spectral density (PSD) of the nonlinear distortions, considered as additive Gaussian noise in absence of nonlinear compensation, is computed. The GNM relies on the hypothesis that large accumulated dispersions scramble the symbols such that according to the central limit theorem, sampled signal's probability density function at the receiver tends to a circular complex Gaussian distribution per polarization, independent of the modulation format; therefore, The domain of validity of GNM is limited as the modulation-dependent contributions are absent, and the low-dispersion regime cannot be modeled. 

The second approach was laid down in a seminal paper by Mecozzi and Essiambre, \cite{Mecozzi2012 : JLT2012}, where the RP method was rigorously applied to the equivalent nonlinear fiber channel comprising the multi-span amplified fiber-optic link and the matched-filtered sampled ideal coherent receiver\footnote {By ideal coherent receiver we mean that timing, frequency offset, source phase noise, and local oscillator phase noise are known to the receiver, and all components are assumed ideal.}. This work was based on the pioneering work of Mecozzi in 2000 \cite{Mecozzi2000 : PTL2000}, where the FWM of optical Gaussian pulses in an optical fiber was rigorously studied for the first time. The main result of this work is computing the third-order time-domain Volterra series coefficients of the equivalent nonlinear fiber channel, which are sufficient to exactly express nonlinear distortions up to the first-order of fiber nonlinear coefficient. Based on \cite{Mecozzi2012 : JLT2012}, a general more accurate theory of nonlinear impairments was developed by Dar \cite{Dar2013 : OptXpress2013, Dar2014 : OptXpress2014, Dar2015 : JLT2015, Dar2016 : JLT2016}, where the impact of modulation format was properly taken into account, and the passage to the high dispersion regime is no more a requirement. He was able to transform the summation over the magnitude square of all Volterra coefficients, which is necessary to obtain the variance of the nonlinear impairments when considered as noise, into equivalent integral representations, and then efficiently compute those integrals by standard Monte Carlo sampling. Following this work, the GNM was upgraded to the enhanced Gaussian noise model (EGN) \cite{Carena2014 : OptXpress2014}. Recently, the full second-order statistics of the nonlinear impairments, \emph{i.e.}, not just the variance, but the whole autocorrelation of the nonlinear distortions was computed\cite{Golani2016 : JLT2016}. 

The third approach, also inspired by\cite{Mecozzi2012 : JLT2012}, as well as\cite{Vannucci2002 : JLT2002}, was suggested by Serena and Bononi \cite{Serena2013 : JLT2013, Serena2015 : JLT2015}, where they proposed to consider the evolution of the time-domain autocorrelation function of the nonlinear distortions along the fiber-optic link. The dependence on modulation formats, and dispersion management could be taken into account. Monte Carlo simulations are necessary to compute the propagation of the autocorrelation function along the link.  
     
All the above-mentioned works only addressed FWM processes among signal waves. They did not account for the FWM between the signal and the co-propagating distributed noise waves injected by the optical amplifiers along the link, sometimes referred to as nonlinear signal-noise interaction (NSNI). Developing a rigorous theory for NSNI is important for at least two reasons: first, it can accurately quantify the amount by which the performance is degraded in various system configurations, (\emph{i.e.}, for different symbol rates, modulation formats, channel spacings, channel counts, number of spans, dispersion maps, fiber types, amplification schemes, \emph{etc.}); second, it can contribute to answering the questions regarding the fundamental limits of performance improvements provided by nonlinear compensation (NLC). 

The two most studied digital NLC algorithms are the digital backpropagation (DBP) \cite{Ip2008 : JLT2008, Fernandez2015 : ECOC2015}, and the less complex, but less accurate perturbation-based NLC (PNLC) \cite{Tao2011: JLT2011}, \cite{Ghazisaeidi2014 : ECOC2014 NLC}, which is based on the theoretical analysis in\cite{Mecozzi2012 : JLT2012, Mecozzi2000 : PTL2000}. The performance of the single-channel NLC, where only the intra-channel nonlinear impairments are partially equalized, either by DBP or by PNLC, is limited by the cross-channel nonlinear interference (NLI)\cite{Dar2017 : JLT2017}. Recently many researchers have investigated multi-channel DBP \cite{Rafique2011 : OptXpress2011, Temprana2015 : OptX2015, Galdino2017 : OptX2017, Liga2014 : OptXpress2014, Lavery2017 : OptXpress2017, Czegledi2017 : OptXpress2017}. In the absence of stochastic fluctuations due to distributed noise, or random birefringence, NLS and Manakov equations have space-reversal symmetry, thus zero-forcing (ZF) full-field equalization by DBP fully compensates the nonlinear fiber channel. On the other hand, the presence of NSNI, and/or polarization effects in the channel breaks the space-reversal symmetry, consequently full-field DBP equalization gain is reduced\footnote{The impact of polarization mode dispersion (PMD) and polarization-dependent loss (PDL) on the performance of single-channel DBP and PNLC is numerically and experimentally investigated in \cite{Fernandez2016 : ECOC2016}. The impact of PMD on multi-channel DBP is studied in\cite{Czegledi2017 : OptXpress2017}. We do not address polarization effects in the present work.}. It is important to assess the achievable signal-to-noise ratio (SNR) improvement due to multi-channel DBP in presence of real irreversible physical phenomena like NSNI and or random polarization effects, in order to determine the fundamental limits of information transmission in optical fibers, and in order to compare multi-channel DBP with other NLC techniques that are currently being investigated, most importantly, optical phase conjugation\cite{Ellis2016 : OFC2016} and nonlinear Fourier transform \cite{Yousefi2014 : TransInfo2014, Aref2016 : ECOC2016}. The theory presented in this work contributes to answering the questions regarding the constraints imposed on multi-channel DBP by NSNI.

The interaction between signal and noise in nonlinear optics has been studied since 1960's\footnote{Please refer to references and discussions in \cite{Serena2016 : JLT2016} for the research work on nonlinear signal-noise processes in the \emph{pre-coherent era}.}. Since 2010, a few authors studied NSNI in the context of modern coherent fiber-optic transmission systems. Ref. \cite{Bononi2010 : OFT2010} presents a complete numerical investigation of various nonlinear impairments in coherent and non-coherent dispersion-managed (DM) systems, with OOK, BPSK and QPSK formats, where, for each scenario the dominant nonlinear impairment is identified. The impact of NSNI on the performance degradation of DBP for 100G coherent PDM-QPSK systems is studied in\cite{Rafique2011 : OptXpress2011} by numerical simulations. Ref.\cite{Foursa2013 : OFC2013} investigates NSNI in 100G coherent systems for both DU and DM systems, both numerically and experimentally, and shows that dispersion management significantly enhances the degrading effect of NSNI. The impact of PMD on NSNI in 100G coherent systems is studied in\cite{Yi2012 : OptXpress2012} by means of numerical simulations. The first theoretical treatment of NSNI for coherent transmission systems is presented in\cite{Beygi2013 : OptXpress2013}, where a discrete channel model is introduced to calculate the impact of NSNI on the performance of single-channel coherent DU systems either without or with DBP, but the derivations are based on many simplifying assumptions. Ref.\cite{Poggiolini : ECOC2014} numerically examines NSNI-induced system reach degradations, which in some cases amount to 15-20\% in DU PDM-QPSK transmissions, and demonstrates that this reach degradation can be explained by a simple phenomenological modification of the GNM, by taking into account signal depletion by noise. The first detailed model of NSNI for coherent systems, is by Serena\cite{Serena2016 : JLT2016}, which is based on \cite{Serena2015 : JLT2015}. The variance of the NSNI is computed by propagating the autocorrelation of nonlinear distortions along the link by means of numerical simulations. DBP can be included in the analysis. Although this approach is general, the numerical computations necessary to calculate the evolution of the autocorrelation functions are still time consuming. 

In this paper, we present a general theory of NSNI for coherent WDM transmission systems, which is exact up to the first-order in fiber nonlinear coefficient, assuming RP, and has the advantage that it is 2-3 orders of magnitude faster than the approach proposed in\cite{Serena2016 : JLT2016}. Although perturbation approximations other than RP have been used to deal with nonlinear effects in the fiber-optic systems\cite{Secondini2016 : JLT2016}, we adopt RP in this work, as the previously discussed analytical treatments of fiber nonlinearity for modeling signal-signal FWM in coherent WDM transmission systems, which are all based on RP,  have been proved to be adequate in practice. Our derivation is based on\cite{Mecozzi2012 : JLT2012}, and the summation technique introduced in \cite{Dar2013 : OptXpress2013, Dar2014 : OptXpress2014, Dar2015 : JLT2015, Dar2016 : JLT2016}. The theory developed here applies to both DM and DU systems, with heterogeneous spans, and both EDFA and Raman amplification schemes. It can deal with chromatic dispersion to all orders; however, the explicit formulas we present in the last sections only include the second-order dispersion, which is to avoid a too cumbersome presentation. The implications of higher order chromatic dispersion terms beyond the second-order is left for future (cf.\cite{Johannisson2013 : JLT2013} for a discussion of the impact of the third-order dispersion on signal-signal nonlinear distortions). We will then use the developed theory to compute the SNR for systems compensated by ideal multi-channel DBP. This computation is useful in providing an estimate on the achievable information rate with ideal ZF DBP compensation. We do not address more sophisticated nonlinear equalization schemes like stochastic DBP\cite{Irukulapati2014 : TransComm2014}. The paper is organized as follows. In section II we present the preliminary materials; the notation is introduced, and the basic equations describing the transmitted signal, channel, and the coherent receiver are written down. In section III we present the first-order regular perturbation theory of WDM propagation, containing both signal-signal and signal-noise FWM contributions. %In section IV numerical results are presented.
Section V concludes the paper. The detailed derivation of the integrals that appear in the expressions of the formulas for the variance of nonlinear distortions, as well as efficient numerical integration technique to compute those integrals are discussed in the appendix.              

\section{Preliminaries}
In this section we give an exact mathematical description of the  multi-span WDM coherent fiber-optic transmission systems, which is the subject of the investigation of this paper. In II.A we define the basic notations and remind some mathematical relations that will be frequently used in the later derivations. In II.B we describe the multi-span single-polarization optically-amplified fiber-optic channel model based on NLS. We assume that noisy erbium-doped fiber amplifiers (EDFA) are placed at the end of each span. We assume heterogeneous spans, with position-dependent dispersion and loss coefficients, so, if necessary, dispersion management, and/or backward Raman amplification can be included in the analysis. The NLS is transformed to a normalized NLS, derived with amplified spontaneous emission (ASE) noise process, and also with the third-order nonlinear term scaled by an effective power profile. The impact of signal depletion by ASE noise is exactly accounted for in deriving the power profile, and the spatio-temporal autocorrelation function of ASE is calculated. In II.C, we describe the ideal matched-filtered coherent receiver, with ideal front-end, and where symbol-by-symbol detection is assumed and no digital signal processing, except for matched filtering, is applied. All the developments in this and the subsequent sections are for perfectly polarized electromagnetic field and NLS. The extension to double polarization and Manakov equation is made in III.H. The notation introduced in this section is faithful to that of \cite{Mecozzi2012 : JLT2012}. 
\subsection{Notations and basic definitions} \label{sec: Notations and basic definitions}
We start by reviewing a few relations and introducing some notational devices that prove to be useful in the sequel.
In this work we are dealing with stochastic processes that are functions of propagation distance $z$ and time $t$. The randomness is due to information symbols and amplifier noise.  Let's denote a sample waveform by $x(z,t)$. The Fourier transform pair is
\begin{equation} \label{eq: FT}
\tilde x\left( {z,\omega } \right) = \mathcal{F} \left[ {x\left( {z,t} \right)} \right] = \int_{ - \infty }^{ + \infty } {dt\exp \left( {i\omega t} \right)x\left( {z,t} \right),} 	
\end{equation}
 \begin{equation} \label{eq: IFT}
	x\left( {z,t} \right) = {\mathcal{F}^{ - 1}}\left[ {\tilde x\left( {z,\omega } \right)} \right] = 
	\frac{1}{{2\pi }}\int_{ - \infty }^{ + \infty } {d\omega\exp \left( { - i\omega t} \right)\tilde x\left( {z,\omega } \right),} 
\end{equation}
where, $\mathcal{F}$ stands for Fourier transform, $\mathcal{F}^{-1}$ stands for inverse Fourier transform, $\omega=2\pi f$ is the angular frequency, and $f$ is the frequency. Throughout this paper, a waveform with a tilde on top  is the Fourier transform of the waveform denoted by the same symbol but without tilde. We have:
\begin{equation}
	\mathcal{F} \left[ {x\left( {z,t - {t_0}} \right)} \right] = \exp \left( { + i\omega {t_0}} \right)\tilde x\left( {z,\omega } \right),
\end{equation}
\begin{equation}
\mathcal{F} \left[ {\exp \left( { - i{\omega _0}t} \right)x\left( {z,t} \right)} \right] = \tilde x\left( {z,\omega  - {\omega _0}} \right),	
\end{equation}
and
\begin{multline}
\mathcal{F} \left[ {\exp \left( { - i{\omega _0}t} \right)x\left( {z,t - {t_0}} \right)} \right] = \\
\exp \left[ {i\left( {\omega  - {\omega _0}} \right){t_0}} \right]\tilde x\left( {z,\omega  - {\omega _0}} \right).	
\end{multline}
In this work we consider only the second-order group velocity dispersion (GVD) for simplicity, but, if necessary, higher order dispersion terms can be included in the analysis without posing any problem. We introduce the following notation for the dispersion operator in frequency domain
\begin{equation} \label{eq: dispersion op. def. fd}
{\hat {\mathcal{D}}_{z'}}\left[ {\tilde x\left( {z,\omega } \right)} \right] = {\exp \left( {i\frac{{{\omega ^2}}}{2}} {\int_0}^{z'} dz''{\beta _2}(z'') \right)}\tilde x\left( {z,\omega } \right),
\end{equation}
and in the time domain
\begin{multline} \label{eq: dispersion op. def. td}
{\hat {\mathcal{D}}_{z'}}\left[ {x\left( {z,t} \right)} \right] = \\
{\mathcal{F} ^{ - 1}}\left[ {\exp \left( {i\frac{{{\omega ^2}}}{2}} {\int_0}^{z'} dz''{\beta _2}(z'')\right)\tilde x\left( {z,\omega } \right)} \right].
\end{multline}
Note that, for simplicity, we use the same notation for the dispersion operator, notwithstanding whether it is applied to a frequency-domain signal, as per (\ref {eq: dispersion op. def. fd}), or to a time-domain signal, as per (\ref {eq: dispersion op. def. td}). Given the context, this should not cause any ambiguity.
We define the following notations for waveforms crosscorrelations in time domain
\begin{equation} \label{eq: xcorr td. def.}
	\left\langle {x\left( {z,t} \right),y\left( {z',t'} \right)} \right\rangle  = \int_{ - \infty }^{ + \infty } {d\tau} {x^*}\left( {z,t+\tau} \right)y\left( {z',t'+\tau} \right),
\end{equation}
and in frequency domain
\begin{equation} \label{eq: xcorr fd. def.}
	\left\langle {\tilde x\left( {z,\omega } \right),\tilde y\left( {z',\omega' } \right)} \right\rangle  = \int_{ - \infty }^{ + \infty } {d\nu} {\tilde x^*}\left( {z,\omega +\nu} \right)\tilde y\left( {z',\omega '+\nu} \right),
\end{equation}
where, the superscript $*$ stands for complex conjugation. 
On the other hand, we use the notation $\left\langle {x\left( {z,t} \right)} \right\rangle$ to denote the ensemble average over the space of all sample waveforms of the stochastic process $x(z,t)$. The randomness of the processes in this work is due to information symbols, which are assumed to be independent identically distributed (i.i.d.) discrete random variables with phase-isotropic distributions, and also due to amplifier noise.

In our notation, the Parseval's theorem is stated as follows
\begin{equation} \label{eq: Parseval theorem}
\left\langle {x\left( {z,t} \right),y\left( {z',t} \right)} \right\rangle  = \frac{1}{{2\pi }}\left\langle {\tilde x\left( {z,\omega } \right),\tilde y\left( {z',\omega } \right)} \right\rangle. 	
\end{equation}
The following \textit {dispersion exchange formula} (DEF), which is easily proven by Parseval's theorem, will be extensively used in this work:
 \begin{equation} \label{eq: DEF def.}
\left\langle {x\left( {z,t} \right),{{\hat {\mathcal{D}}}_{z'}}\left[ {y\left( {z,t} \right)} \right]} \right\rangle  = \left\langle {{{{\hat {\mathcal{D}}}^\dag}_{z'}}\left[ {x\left( {z,t} \right)} \right],y\left( {z,t} \right)} \right\rangle, 	
\end{equation}
where, $\hat{\mathcal D}^{\dag}_{z'}$ is the adjoint of the dispersion operator $\hat{\mathcal{D}}_{z'}$, which is defined to be  
\begin{equation} \label{eq: dispersion op. def. fd adjoint}
{{\hat {\mathcal{D}}^\dag}_{z'}}\left[ {\tilde x\left( {z,\omega } \right)} \right] = {\exp \left( -{i\frac{{{\omega ^2}}}{2}} {\int_{0}}^{z'} dz''{\beta _2}(z'') \right)}\tilde x\left( {z,\omega } \right).
\end{equation}

All waveforms are assumed to be base-band analytical signals. We use the first subscript for continuous waveforms to denote time-shifts by multiples of symbol duration, \emph{i.e.},
\begin{equation}
	{x_k}\left( {z,t} \right) = x\left( {z,t - {T_k}} \right),	
\end{equation}
where, $T_k = kT$, $T$ is the symbol duration and $k$ is an arbitrary integer. When used to decorate a discrete random variable, the first subscript $k$ denotes the $k$'th symbol. The second subscript, both for continuous waveforms, and for discrete random variables, denotes the WDM channel index. Thus $a_{k, s}$ denotes the $k$'th symbol of the $s$'th channel, and $x_{k, s}(z,t)$ denotes the base-band waveform of the $s$'th channel time shifted by $kT$. The channel of interest (COI) is indexed $s=0$. If the second index is zero it can be optionally dropped in order to simplify the notation, therefore: $a_{k} = a_{k, 0}$, and $x_k(z,t)=x_{k,0}(z,t)$. The total optical field at distance $z$ and time $t$ is denoted by $E(z,t)$. The total optical field at fiber input ($z=0$) is written as
\begin{multline} \label{eq: E(0,t) def.}
E\left( {0,t} \right) = \sum\limits_k {{a_k}{A_k}\left( {0,t} \right)}  + \\
\sum\limits_{k,s \ne 0} {{a_{k,s}}{A_{k,s}}\left( {0,t - \delta {T_s}} \right)\exp \left[ { - i{\Omega _s}t + i{\phi _s}\left( 0 \right)} \right]}, 	
\end{multline}
where, as mentioned above, $a_k$ is the $k$'th symbol of the COI, $a_{k,s}$ is the $k$'th symbol of the $s$'th adjacent channel, $\Omega_s$ is the center frequency detuning of the $s$'th adjacent channel with respect to COI, $\delta T_s$ is the time offset of the $s$'th adjacent channel with respect to COI, $\phi_s(0)$ is the initial, \emph{i.e.}, $z=0$, phase offset of the $s$'th adjacent channel with respect to COI, $A_0(0,t)$ is the pulseshape of the COI, and $A_{0,s}(0,t)$ is the pulseshape of the $s$'th channel. In this work, we suppose that $A_{0,s}(0,t) = A_0(0,t)$ for all $s$.
The energy, $\mathcal E$, of the pulseshapes is given by the following integral
\begin{equation}
\mathcal{E} = \int_{ - \infty }^{ + \infty } {{{\left| {{A_0}\left( {0,t} \right)} \right|}^2}dt}. 	
\end{equation}
The normalized pulseshape is defined as
\begin{equation} \label{eq: U0(0,t) def.}
{U_0}\left( {0,t} \right) = \frac{{{A_0}\left( {0,t} \right)}}{{\sqrt {\mathcal{E}} }}.
\end{equation}
In this work we w assume Nyquist pulse-shaping is applied to all channels. As a consequence, the following orthogonality relation holds for the pulses
\begin{equation}
\left\langle {{A_k}\left( {0,t} \right),{A_h}\left( {0,t} \right)} \right\rangle  = {\mathcal{E}}{\delta _{hk}}, 
\end{equation}
which is equivalent to the following orthonormality condition for the normalized pulses
\begin{equation} \label{eq: normalized orthonormality relation}
\left\langle {{U_k}\left( {0,t} \right),{U_h}\left( {0,t} \right)} \right\rangle  = {\delta _{hk}},	
\end{equation}
where, $\delta _{hk}$ is the Kronecker delta function.
\subsection{Nonlinear Schr\"odinger equation (NLS)}
If polarization effects are ignored, the total optical field satisfies the the NLS, \emph{i.e.}, 
\begin{equation} \label {eq: NLS}
{\partial _z}E = \frac{{g\left( z \right) - \alpha(z) }}{2}E - i\frac{{{\beta _2(z)}}}{2}\partial _t^2E + i\gamma(z) {\left| E \right|^2}E + n\left( {z,t} \right),	
\end{equation}
where, $g(z)$ is the local power gain coefficient, $\alpha(z)$ is the local fiber attenuation coefficient, $\beta_2(z)$ is the local GVD coefficient, $\gamma(z)$ is the local fiber nonlinear Kerr coefficient, and $n(z,t)$ is the amplified spontaneous emission (ASE) noise source.
Let's denote the COI wavelength by $\lambda_0$. The measured channel carrier angular frequency is $\omega_0 = 2\pi c/\lambda_0$, where $c$ is the speed of light. Note that, $\gamma = 2\pi n_2/\lambda_0 A_{eff}$, where $n_2$ is the fiber nonlinear Kerr refractive index, and $A_{eff}$ is the fiber effective area.

In writing (\ref {eq: NLS}) we assumed that the local power gain coefficient is frequency-independent, \emph{i.e.},
\begin{equation}
	 g\left( {z,{\omega }} \right) =  g\left( {z,{\omega _0}} \right) = g\left( z \right).
\end{equation}
The frequency-independent local power gain coefficient can be explicitly written as
\begin{equation} \label{eq: g(z) definition}
g\left( z \right) = \sum\limits_{n = 1}^{{N_s}} {{g_n}\delta \left( {z - {z_{n}}} \right)}, 
\end{equation}
where, $\delta(\cdot)$ is the Dirac delta function, $g_n$ is the local gain coefficient of the EDFA placed at the end of the $n$'th span, $N_s$ is the total number of spans, and $z_n$ is the coordinate of the end of the $n$'th span. We also define
\begin{equation} \label{eq: z_0 zero}
z_0=0.
\end{equation}
The ASE noise source in (\ref {eq: NLS}) is a complex circular Gaussian random process with zero mean and the following time-domain autocorrelation function
\begin{equation}
\left\langle {{n^*}\left( {z,t} \right)n\left( {z',t'} \right)} \right\rangle  = C\left( {z,t - t'} \right)\delta \left( {z - z'} \right).	
\end{equation}
The Fourier transform of the noise source autocorrelation, \emph{i.e.}, the space-dependent power spectral density of the ASE source is
\begin{equation}
	\left\langle {{{\tilde n}^*}\left( {z,\omega } \right)\tilde n\left( {z',\omega '} \right)} \right\rangle  = 2\pi \tilde C\left( {z,\omega } \right)\delta \left( {z - z'} \right)\delta \left( {\omega  - \omega '} \right),
\end{equation}
where, we have
\begin{equation} \label{eq: Csw noise } 
\tilde C(z,\omega) = \hbar\omega _0\sum\limits_{n = 1}^{{N_s}} {{g_n}{n_{\text{sp},n}}(\omega)\delta(z-z_n)},
\end{equation}
where, $n_{\text{sp},n}(\omega)$ is the noise figure of the $n$'th amplifier, and $\hbar$ is Planck's constant divided by $2\pi$. 

Now, in order to simplify further developments, we derive a normalized version of NLS by factoring out the power profile. Let's define the normalized total optical field $U(z,t)$ as follows
\begin{equation} \label{eq: U definition}
	E\left( {z,t} \right) = \Psi \left( z \right)U\left( {z,t} \right),
\end{equation}
where, the power envelop function $\Psi(z)$ satisfies the following equation by definition
\begin{equation} \label{eq: psi ODE}
	\frac{d}{{dz}}\Psi \left( z \right) = \frac{{g\left( z \right) - \alpha(z)}}{2}\Psi \left( z \right),
\end{equation}
with the initial condition
\begin{equation} \label{eq: psi_z init. cond.}
\Psi \left( 0 \right) = \sqrt {\mathcal{E}}. 
\end{equation}
Now, we define the normalized power profile function $f(z)$ as
\begin{equation} \label{eq: f(z) intial definition }
f(z) = \frac{1}{\sqrt\mathcal E}\Psi(z).
\end{equation}
We can solve (\ref{eq: psi ODE}) and apply the initial condition (\ref{eq: psi_z init. cond.}) and the definition (\ref{eq: f(z) intial definition }) to obtain
\begin{equation} \label{eq: f explicit def.} 
f\left( z \right) = \exp \left\{ {\int_0^z {dz'\left[ {g\left( {z'} \right) - \alpha \left( {z'} \right)} \right]} } \right\}.
\end{equation}
After substituting (\ref{eq: U definition}) into (\ref {eq: NLS}) and using (\ref{eq: psi ODE}) and (\ref{eq: f(z) intial definition }), we can derive the following normalized NLS
\begin{equation} \label{eq: Normalized NLS}
{\partial _z}U =  - i\frac{{{\beta _2(z)}}}{2}\partial _t^2U + i{\mathcal{E}}\gamma(z)f\left( z \right){\left| U \right|^2}U + \frac{{n\left( {z,t} \right)}}{{\sqrt {{\mathcal{E}}f\left( z \right)} }},
\end{equation}

In this work, for simplicity, we assume that all channels have the same average launch power per channel, which is denoted by $P$; we have 
\begin{equation} \label{eq: P_0 definition}
{P} = \frac{\mathcal{E}}{T};
\end{equation}
Moreover, we assume that all EDFAs operate in the constant output power mode. Let's denote the ASE power at the output of the $n^{\text {th}}$ EDFA, \emph{i.e.}, the EDFA placed at the end of the $n^{\text {th}}$ span, by $N_n$, and the optical power transferred from the $(n-1)^{\text {th}}$ span to the $n^{\text {th}}$ span, after amplification by the $n^{\text {th}}$ EDFA, by $S_n$. We have
\begin{equation} \label{eq: N_n definition}
\begin{array}{*{20}{c}}
{N_n= (e^{g_{n}}-1)\sigma_{n}^2}&{n = 1, \ldots ,{N_s}},
\end{array}
\end{equation}
where
\begin{equation} \label{eq: sig2_n definition}
\sigma _n^2 = \hbar \omega_0\int_{ - \infty }^{ + \infty } {d\omega n_{\text{sp},n}}(\omega).
\end{equation}
Note that $N_0=0$, and $S_0=P$. We have
 \begin{equation} \label{eq: S_n definition}
\begin{array}{*{20}{c}}
{S_n=e^{g_{n-1}-l_{n-1}}(S_{n-1}+N_{n-1})}&{n = 1, \ldots ,{N_s}},
\end{array}
\end{equation}
where, the $l_n$ is the loss exponent of the $n^{\text {th}}$ span, which is 
\begin{equation} \label{eq: loss exponent}
{l_n} = \int_{{z_{n - 1}}}^{{z_n}} {dz\alpha \left( z \right)}.
\end{equation}
 The constant output power assumption for EDFAs amounts to
\begin{equation} \label{eq: Constant EDFA output power mode condition}
\begin{array}{*{20}{c}}
{{S_n} + {N_n} = {P}}&{n = 1, \ldots ,{N_s}}.
\end{array}
\end{equation}
Note that the noise figure function $n_{\text{sp},n}(\omega)$ is dimensionless; therefore, the noise variance $\sigma_n^2$ in (\ref {eq: sig2_n definition}) has the dimension of power. Now, let's define the following dimensionless parameter
\begin{equation} \label{eq: zeta definition}
{\zeta _n} = \frac{{\sigma _n^2}}{{{P}}},
\end{equation}
as well as
\begin{equation} \label{eq: depletion exponent definition}
{d_n} = {l_n} - {g_n},
\end{equation}
where, the $d_n$ is the signal gain depletion exponent in the $n^{\text {th}}$ by the ASE power generated at that span\footnote{The is different from signal depletion by NSNI, which is discussed in \cite{Poggiolini : ECOC2014}. Here all nonlinear distortions, including NSNI, contribute to $S_n$.}. Using (\ref{eq: N_n definition}), (\ref{eq: S_n definition}), (\ref {eq: Constant EDFA output power mode condition}), (\ref {eq: zeta definition}), and (\ref {eq: depletion exponent definition}), we have
\begin{equation} \label{eq: d_n formula}
d_n = \ln(\frac{1+e^{l_n}\zeta_n}{1+\zeta_n}),
\end{equation}
where $n=1,\ldots,N_s$. We also define
\begin{equation} \label{eq: d_0 definition}
d_0=0.
\end{equation}
The accumulated signal depletion exponent is denoted by $\delta_n$, and is given by
\begin{equation} \label{eq: delta_n definition}
{\delta _n} = \sum\limits_{m = 0}^{n} {{d_m}}.
\end{equation}
Using these definitions we find the following expression for the normalized power profile
\begin{equation} \label{eq:f(z) explicit formula}
f\left( z \right) = \begin{array}{*{20}{c}}
{\exp \left[ { - {\delta _{n-1}} - \int_{{z_{n - 1}}}^z {dz'\alpha \left( {z'} \right)} } \right],}&{{z_{n - 1}} \le z < {z_n}}, 
\end{array}
\end{equation}
for ${n = 1, \ldots ,{N_s}}$. This expression for the normalized power profile is general, in that spans with arbitrary length and loss coefficients can be modeled, the noise figure of the EDFAs are different; moreover, backward Raman amplification can be modeled by properly defining a $z$-dependent loss coefficient function $\alpha(z)$. Finally note that, in order to model the system when the EDFAs operate in the constant gain mode, we only need to force $d_n=0$ for $n=0,\ldots,N_s$.
\subsection{Coherent Receiver}
We assume ideal matched-filter symbol-by-symbol coherent receiver. The total transmission distance is denoted by $L$. The impulse response of the matched filter is denoted by $u_{f}(t)$. We have
\begin{equation} \label{eq: uf in terms of Dt}
{u_f}\left( t \right) = {{D_0}^*}\left( { - t} \right),
\end{equation}
where,
\begin{equation} \label{eq:Dt pulse definition}
{D_0}\left( t \right) = {{\hat {\mathcal D}}_L}\left[ {{U_0}\left( {0,t} \right)} \right].
\end{equation}
\footnote{\emph{i.e.}, we assume that the digital signal processing in the coherent receiver only consists of dispersion compensation and linear matched filtering by a filter whose impulse response is the Nyquist pulse.}The photocurrent at the output of the matched filter is denoted by ${\mathcal I}$, and is given by the following equation
\begin{equation} \label{eq: I definition}
{\mathcal I}\left( t \right) = {u_f}\left( t \right) \otimes U\left( {L,t} \right),
\end{equation}
where, the symbol $\otimes$ stands for the convolution operation in time domain.
The sampled photocurrent at time $t = kT$ is
\begin{equation} \label{eq: I_k definition}
{\mathcal I}_k = {\mathcal I}\left( {kT} \right) = \int_{ - \infty }^{ + \infty } {dt'D_0^*\left( {t' - kT} \right)U\left( {L,t'} \right)}.
\end{equation}
Using the notation introduced in (\ref{eq: xcorr td. def.}) and (\ref{eq:Dt pulse definition}), the (\ref {eq: I_k definition}) can be rewritten as
\begin{equation} \label{eq: I_k xcorr}
{\mathcal I}_k = \left\langle {{{\hat {\mathcal D}}_L}\left[ {{U_k}\left( {0,t} \right)} \right],U\left( {L,t} \right)} \right\rangle.
\end{equation}
\section{First-order regular perturbation}
In this section, we build upon the material developed in the previous section to lay down the complete first-order RP treatment of the multi-span coherent WDM systems, including signal-signal and signal-noise FWM terms. The general formulation of RP is presented in III.A. The zeroth order, (or linear), solution is presented in III.B. This solution is composed of dispersed signal terms and additive ASE noise terms. In III.C we express the ASE term appearing in the zeroth order solution as a Karhunen-Lo\`eve expansion series in the signal basis. This step is crucial for the further development of the theory. In III.D the first-order RP solution containing both signal-signal and signal-noise FWM contributions to the sampled photo-current at the receiver side is derived. In III.E the variance of the ASE noise is calculated. In III.F the variance of the nonlinear signal-signal distortions is calculated. In III.G the variance of the NSNI distortions is calculated. In III.H the results are extended to the double-polarization case assuming the physics is governed by the Manakov equation. The variance of signal-signal and signal-noise distortions are expressed as sums over the so-called $\mathcal{X}$- and $\chi-$coefficients. These coefficients are represented as multi-dimensional integrals, and have to be numerically computed by Monte Carlo integrations. The detailed derivation of $\mathcal{X}$- and $\chi-$coefficients is the subject of the appendix.  
\subsection{General formulation}
From now on we suppose that the fiber nonlinear coefficient $\gamma$ is not a function of $z$\footnote{If necessary, the $f(z)$ and the $\tilde C(z,\omega)$ can be redefined in order to account for the $z$-dependence of $\gamma$, cf. (\ref{eq: Csw noise }), (\ref{eq: f explicit def.}), and (\ref{eq: Normalized NLS}).}. Let's write the total normalized optical field as a regular perturbation series with respect to the fiber nonlinear coefficient $\gamma$
\begin{equation} \label {eq: Perturbation series}
U\left( {z,t} \right) = \sum\limits_{n = 0}^\infty  {{\gamma ^n}{u^{\left( n \right)}}\left( {z,t} \right)},
\end{equation}
where, the ${{u^{\left( n \right)}}\left( {z,t} \right)}$ is the $n$'th order perturbation correction to the total normalized optical field $U(z,t)$. Let's denote the regular perturbation approximation of $U(z,t)$, when only zeroth order and first order terms are kept in the expansion (\ref{eq: Perturbation series}), by $U_{FRP}(z,t)$. We have
\begin{equation} \label{eq: first-order perturbation}
U_{FRP}\left( {z,t} \right) = {u^{\left( 0 \right)}}\left( {z,t} \right) + \gamma {u^{\left( 1 \right)}}\left( {z,t} \right).	
\end{equation}
Note that
\begin{equation} \label{eq: U vs. U_FRP}
U\left( {z,t} \right) = {U_{FRP}}\left( {z,t} \right) + {\mathcal O}\left( {{\gamma ^2}} \right).
\end{equation}
In order to calculate the $U_{FRP}(z,t)$ as per (\ref {eq: first-order perturbation}), we need to find $u^{(0)}(z,t)$ and $u^{(1)}(z,t)$. To do so, we substitute (\ref {eq: Perturbation series}) into (\ref {eq: Normalized NLS}), and separate the zeroth order and the first order terms. For the zeroth order term we obtain 
\begin{equation} \label{eq: zeroth-order pde}
{\partial _z}{u^{\left( 0 \right)}}\left( {z,t} \right) =  - i\frac{{{\beta _2}}(z)}{2}\partial _t^2{u^{\left( 0 \right)}}\left( {z,t} \right) + \frac{{n\left( {z,t} \right)}}{{\sqrt {{\mathcal{E}}f\left( z \right)} }},
\end{equation}
and for the first-order we obtain
\begin{multline}  \label{eq: first-order pde}
{\partial _z}{u^{\left( 1 \right)}}\left( {z,t} \right) =  - i\frac{{{\beta _2}}(z)}{2}\partial _t^2{u^{\left( 1 \right)}}\left( {z,t} \right) + \\
i{\mathcal{E}} f\left( z \right){\left| {{u^{\left( 0 \right)}}\left( {z,t} \right)} \right|^2}{u^{\left( 0 \right)}}\left( {z,t} \right).
\end{multline}
\subsection{Zeroth-order solution}
The initial condition for the zeroth-order equation, (\ref{eq: zeroth-order pde}), is
\begin{equation}
{u^{\left( 0 \right)}}\left( {0,t} \right) = s\left( t \right) + \frac{{n\left( {0,t} \right)}}{{\sqrt {\mathcal{E}} }} = s\left( t \right),
\end{equation}
where, we assumed the optical signal-to-noise ratio (OSNR) at the transmitter side is infinite, therefore $n(0,t)=0$. The waveform $s(t)$ is the total WDM signal injected into the fiber channel at the transmitter side. It can be explicitly written as 
\begin{multline} \label{eq: s definition}
s\left( t \right) = \sum\limits_k {{a_k}u_k^{\left( 0 \right)}\left( {0,t} \right)}  + \\
\sum\limits_{k,s \ne 0} {{a_{k,s}}u_{k,s}^{\left( 0 \right)}\left( {0,t - \delta {T_s}} \right)\exp \left[ { - i{\Omega _s}t + i{\phi _s}\left( 0 \right)} \right]},
\end{multline}
where, the noramlized base-band pulses for all channels are assumed to be the Nyquist pulses, \emph{i.e.},
\begin{equation} \label{eq: sinc definition}
u_0^{\left( 0 \right)}\left( {0,t} \right) = u_{0,s}^{\left( 0 \right)}\left( {0,t} \right) = {\frac{1}{\sqrt T}}{\rm{sinc}}\left( {\frac{t}{T}} \right), 
\end{equation}
where, 
\begin{equation}
{\rm{sinc}}\left( x \right) = \frac{{\sin \left( {\pi x} \right)}}{{\pi x}}.
\end{equation}
The Fourier transform of (\ref {eq: sinc definition}) is 
\begin{equation} \label{eq: sinc spectrum}
\tilde u_0^{\left( 0 \right)}\left( {0,\omega } \right) = \left\{ {\begin{array}{*{20}{c}}
{\begin{array}{*{20}{c}}
{\sqrt T }&{\left| \omega  \right| < \frac{\pi }{T}}
\end{array}}\\
{\begin{array}{*{20}{c}}
0&{\left| \omega  \right| \ge \frac{\pi }{T}}
\end{array}}
\end{array}} \right..
\end{equation}
As mentioned previously, the information symbols in all channels are assumed to be i.i.d. random variables. Furthermore, given the normalization conventions adopted in this work, their second moment is equal to unity, \emph{i.e.}, 
\begin{equation} \label{eq: symbol's second order moment normalization}
\left\langle {{{\left| {{a_k}} \right|}^2}} \right\rangle  = \left\langle {{{\left| {{a_{k,s}}} \right|}^2}} \right\rangle  = 1.
\end{equation}
The solution of the zeroth order equation (\ref {eq: zeroth-order pde}) is
\begin{equation} \label{eq: zeroth order solution}
u^{(0)}({z,t}) = {{\hat {\mathcal{D}}}_z} [ s(t) 	+ {{u'}_{{\rm{ASE}}}}({z,t} ) ].
\end{equation}
This zeroth order solution is the sum of dispersed transmitted pulses and the additive ASE field, which is 
\begin{equation}
{{u'}_{{\rm{ASE}}}}\left( {z,t} \right) = \int_0^z {dz'} {\hat {\mathcal D}_{z'}}^\dag\left[ {\frac{{n\left( {z',t} \right)}}{{\sqrt {{\mathcal{E}}f\left( {z'} \right)} }}} \right].
\end{equation}
The total ASE field is a complex circular Gaussian random process, with zero mean and the following frequency-domain autocorrelation function
\begin{equation} \label{eq: ASE autocorrelation}
\left\langle {\tilde {u'}_{ASE}^*({z,\omega})\tilde u'_{ASE}({z',\omega'})} \right\rangle  = 2\pi \kappa({z,z';\omega})\delta({\omega - \omega'}),
\end{equation}
where, 
\begin{equation} \label{eq: kappa definition}
\kappa \left( {z,z';\omega } \right) = \int_0^{\min \left( {z,z'} \right)} {dz''} \frac{{\tilde C\left( {z'',\omega } \right)}}{{{\mathcal{E}}f\left( {z''} \right)}}.
\end{equation}
For later convenience, we define
\begin{equation} \label{eq: xi definition}
\xi({z;\omega}) = \frac{1}{\hbar\omega_0}\int_0^{z}{dz''} \frac{{\tilde C( {z'',\omega })}}{{f({z''})}};
\end{equation}
therefore, the normalized PSD of the ASE field, $\kappa$, can be rewritten as
\begin{equation} \label{eq: relation of kappa with xi}
\kappa(z,z';\omega)=\frac{\hbar\omega_0}{\mathcal{E}}\xi(\min(z,z');\omega).
\end{equation}
After substituting (\ref{eq: Csw noise }) and (\ref{eq:f(z) explicit formula}) into (\ref {eq: xi definition}), and carrying out the integration, we obtain
\begin{equation} \label {eq: xi detailed formula}
\xi(z,\omega) = \sum\limits_{n = 1}^{N_s} {g_n e^{\delta_n}n_{\text{sp},n}(\omega)\theta(z-z_n)},	
\end{equation}
where, the function $\theta(\cdot)$ is the Heaviside step function, \emph{i.e.},

\begin{equation}\label{eq: Heaviside step func. def.}
\theta \left( x \right) = \left\{ {\begin{array}{*{20}{c}}
  0&{x \geq 0} \\ 
  1&{x < 0} 
\end{array}} \right..
\end{equation}
For later convenience, let's define ${\xi _n}\left( \omega  \right) = {g_n}{e^{{\delta _n}}}{n_{{\rm{sp}},n}}\left( \omega  \right)$, for $n = 1, \ldots ,{N_s}$, and define ${\xi _0}\left( \omega  \right) = 0$. Using this notation, the (\ref {eq: xi detailed formula}) can be written as
\begin{equation} \label{eq: xi(z,omega) simplified formula}
\xi \left( {z,\omega } \right) = \sum\limits_{n = 0}^{{N_s}} {{\xi _n}\left( \omega  \right)\theta \left( {z - {z_n}} \right)}.
\end{equation}
Note from (\ref {eq: kappa definition}) that the autocorrelation of ${{u'}_{{\rm{ASE}}}}$ is independent of dispersion. This is normal, since the all-phase linear filtering of a complex circular Gaussian process does not change its second-order statistical properties. We can therefore replace ${{{u'}_{{\rm{ASE}}}}\left( {z,t} \right)}$ in (\ref {eq: zeroth order solution}) by an equivalent ASE field ${u_{{\rm{ASE}}}}\left( {z,t} \right)$, defined as
\begin{equation} \label{eq: u_ASE def.}
{u_{{\rm{ASE}}}}\left( {z,t} \right) = \int_0^z {dz'} \frac{{n\left( {z',t} \right)}}{{\sqrt {{\mathcal{E}}f\left( {z'} \right)} }}.
\end{equation}
Note that the ${u_{{\rm{ASE}}}}\left( {z,t} \right)$ is also a complex circular Gaussian random process with zero mean and the same autocorrelation function as that of ${u'_{{\rm{ASE}}}}\left( {z,t} \right)$, \emph{i.e.},
\begin{equation} \label{eq: ASE autocorrelation}
\left\langle {\tilde u_{ASE}^*\left( {z,\omega } \right)\tilde u_{ASE}^{}\left( {z',\omega '} \right)} \right\rangle  = 2\pi \kappa \left( {z,z';\omega } \right)\delta \left( {\omega  - \omega '} \right).
\end{equation}
From now on, instead of (\ref {eq: zeroth order solution}) we use the following equation  
\begin{equation} \label{eq: zeorth-order complete solution}
u^{(0)}({z,t}) = {{\hat {\mathcal{D}}}_z}[s(t) + {u_{{\rm{ASE}}}}({z,t})].
\end{equation}
Using (\ref {eq: s definition}), equation (\ref {eq: zeorth-order complete solution}) can be developed as
\begin{multline} \label{eq: zeorth-order complete solution detailed}
{u^{\left( 0 \right)}}\left( {z,t} \right) = \sum\limits_k {{a_k}u_k^{\left( 0 \right)}\left( {z,t} \right)}  + \\
 \sum\limits_{k,s \ne 0} {{a_{k,s}}u_{k,s}^{\left( 0 \right)}\left( {z,t - t_s(z)} \right)
\exp \left[ { - i{\Omega _s}t + i{\phi _s}\left( z \right) } \right]}  + \\
{{\hat {\mathcal{D}}}_z}\left[ {{u_{{\rm{ASE}}}}\left( {z,t} \right)} \right],
\end{multline}
where, the zeroth order dispersed pulse of the $k$'th symbol of the $s$'th channel after propagating up to distance $z$ is
\begin{equation} \label{eq:}
u_{k,s}^{\left( 0 \right)}\left( {z,t} \right) = {\hat {\mathcal D}_z}\left[ {u_{k,s}^{\left( 0 \right)}\left( {0,t} \right)} \right].
\end{equation}
The walk-off time shift between COI and the $s$'th channel is
\begin{equation} \label{eq: ts(z) def.}
{t_s}\left( z \right) = \delta {T_s} + {\Omega _s}\int_0^z {dz'{\beta _2}\left( {z'} \right)}.
\end{equation}
The relative phase shift between COI and the $s$'th channel is
\begin{equation} \label{eq: phi_s(z) def.}
{\phi _s}\left( z \right) = {\phi _s}\left( 0 \right) + \frac{{\Omega _s^2}}{2}\int_0^z {dz'{\beta _2}\left( {z'} \right)}.
\end{equation}
In the rest of this work we suppose that $\delta T_s=0$, and that the WDM channels are uniformly spaced, with channel spacing equal to $\Delta \Omega$; therefore, $\Omega_s = s\Delta \Omega$, and $\Omega_{s+s'}= \Omega_s+\Omega_{s'}$.
\subsection{Karhunen-Lo\`eve (KL) expansion of the ASE field}
In order to simplify modeling the signal-noise interaction in the next section, we can put signal and noise on equal footing. To do so, we consider the Karhunen-Lo\`eve (KL) expansion, \cite{Papoulis2002}, of the ASE term in (\ref {eq: zeorth-order complete solution detailed}), on the basis of the zeroth order dispersed Nyquist pulses as follows
\begin{multline} \label{eq: ASE KL expansion}
{{\hat {\mathcal{D}}}_z}\left[ {{u_{{\rm{ASE}}}}\left( {z,t} \right)} \right] = \sum\limits_k {{w_k}\left( z \right)u_k^{\left( 0 \right)}\left( {z,t} \right)}  \\
+ \sum\limits_{k,s \ne 0} {{w_{k,s}}\left( z \right)u_{k,s}^{\left( 0 \right)}\left(z,t - t_s(z)\right)
\exp \left[ { - i{\Omega _s}t + i{\phi _s}\left(z\right)} \right]},
\end{multline}
where, the $w_k(z)$ and $w_{k,s}(z)$ are $z$-dependent random variables, which are the KL expansion coefficients of the ASE field, in the orthonormal basis of the time-shifted WDM Nyquist pulses. Note that, we have ignored the ASE degrees of freedom that fall outside the signal band. This approximation is justified by numerical simulations in \cite{Serena2016 : JLT2016}. The KL expansion coefficients of the ASE field of the COI are explicitly written as
\begin{equation} \label{eq: wz definition}
{w_k}\left( z \right) = \left\langle {u_k^{\left( 0 \right)}\left( {z,t} \right),{{\hat {\mathcal{D}}}_z}\left[ {{u_{{\rm{ASE}}}}\left( {z,t} \right)} \right]} \right\rangle;
\end{equation}
similarly, the KL expansion coefficients of the ASE field added to the $s$'th adjacent channel signal are
\begin{multline} \label{eq: wzs definition}
w_{k,s}(z) = \langle u_{k,s}^{(0)}(z,t - t_s(z))\exp [-i\Omega_st + i\phi_s(z)] \\
 {,\hat {\mathcal D}_z[u_{\rm{ASE}}(z,t)]} \rangle .
\end{multline}
We can substitute (\ref {eq: ASE KL expansion}) into (\ref {eq: zeorth-order complete solution detailed}) to obtain
\begin{multline} \label {eq: zeorth-order complete solution detailed with KL}
{u^{\left( 0 \right)}}\left( {z,t} \right) = \sum\limits_k {\left( {{a_k} + {w_k}} \right)u_k^{\left( 0 \right)}\left( {z,t} \right)}  + \\ \sum\limits_{k,s \ne 0} {\left( {{a_{k,s}} + {w_{k,s}}} \right)u_{k,s}^{\left( 0 \right)}\left( {z,t - t_s} \right)
\exp \left[ { - i{\Omega _s}t + i{\phi _s}} \right]},	
\end{multline}
where, in order to simplify the notation, we have dropped the $z$-dependence of $w_k(z)$, $w_{k,s}(z)$, $t_s(z)$, and $\phi_s(z)$ in writing (\ref{eq: zeorth-order complete solution detailed with KL}).
Using the DEF, as per (\ref {eq: DEF def.}), (\ref {eq: wz definition}) can be rewritten as
\begin{equation} \label{eq: wz definition, disp. exchanged}
{w_k}\left( z \right) = \left\langle {u_k^{\left( 0 \right)}\left( {0,t} \right),{u_{{\rm{ASE}}}}\left( {z,t} \right)} \right\rangle;
\end{equation}
similarly, (\ref {eq: wzs definition}) can be written as
\begin{multline} \label{eq: wzs definition, disp. exchanged}
w_{k,s}(z) = \langle u_{k,s}^{(0)}(0,t - \delta {T_s})
\exp [-i\Omega_st + i\phi_s(0)] \\
,u_{\rm{ASE}}(z,t) \rangle.
\end{multline}
The KL expansion coefficients ${w_k}\left( z \right)$ and ${w_{k,s}}\left( z \right)$ are zero-mean complex circular Gaussian random variables. The $w_{k}(z)$ is independent from  $w_{k,s}(z)$, which is independent from  $w_{k,s'}(z)$ if $s\neq s'$.
The following expressions can be derived for the autocorrelations of the KL expansions of the ASE field added to the COI signals
\begin{equation} \label{eq: w_k autocorr}
\left\langle {w_k^*\left( {z'} \right){w_k}\left( z \right)} \right\rangle  = \frac{1}{{2\pi }}\int_{ - \infty }^{ + \infty } {{{d\omega\left| {{{\tilde u}^{\left( 0 \right)}}\left( {0,\omega } \right)} \right|}^2}\kappa \left( {z,z';\omega } \right)}.
\end{equation}
For the ASE field added to the $s$'th adjacaent channel, we have
\begin{multline} \label{eq: w_k,s autocorr}
\left\langle {w_{k,s}^*\left( {z'} \right){w_{k,s}}\left( z \right)} \right\rangle  = \\
\frac{1}{{2\pi }}\int_{ - \infty }^{ + \infty } {{{d\omega\left| {{{\tilde u}^{\left( 0 \right)}}\left( {0,\omega  - {\Omega _s}} \right)} \right|}^2}\kappa \left( {z,z';\omega } \right)}.
\end{multline}
Assuming that the ASE spectrum is flat over signal bandwidth, we can approximate the above autocorrelation functions as follows: for the COI we have 
\begin{equation} \label{eq: w_k autocorr aprrox} 
\left\langle {w_k^*\left( {z'} \right){w_k}\left( z \right)} \right\rangle  \sim \kappa \left( {z,z';0} \right), 
\end{equation}
and for the $s$'th adjacent channel we have
\begin{equation} \label{eq: w_k,s autocorr approx}
\left\langle {w_{k,s}^*\left( {z'} \right){w_{k,s}}\left( z \right)} \right\rangle  \sim \kappa \left( {z,z';{\Omega _s}} \right).
\end{equation}
\subsection{First-order solution}
The boundary condition for the first-order solution at $z=0$ is
\begin{equation} \label{eq: u1 at z=0}
u^{(1)}(0,t)=0.
\end{equation}
Having solved the zeroth order equation, we now use (\ref {eq: first-order pde}) together with (\ref {eq: u1 at z=0}) to find the first-order regular perturbation correction to the normalized optical field as follows
\begin{multline} \label{eq: first-order solution}
{u^{\left( 1 \right)}}\left( {z,t} \right) = \\
i{\mathcal{E}}{\hat {\mathcal D}_{z}}{\int_0^z {dz'} {f\left( {z'} \right)}{\hat {\mathcal D}_{z'}}^\dag\left[ {{{\left| {{u^{\left( 0 \right)}}\left( {z',t} \right)} \right|}^2}{u^{\left( 0 \right)}}\left( {z',t} \right)} \right]}.
\end{multline}
We denote the first-order RP approximation to the sampled photocurrent ${\mathcal I}_k$ by ${\mathcal J}_k$. using (\ref {eq: I_k xcorr}) and (\ref {eq: first-order perturbation}) we have
\begin{equation} \label{eq: J_k xcorr}
{\mathcal J}_k = \left\langle {{{\hat {\mathcal D}}_L}\left[ {{U_k}\left( {0,t} \right)} \right],{U_{FRP}}\left( {L,t} \right)} \right\rangle.
\end{equation}
Note that due to (\ref{eq: U vs. U_FRP}) we have
\begin{equation} \label{eq: I_k vs. J_k}
{\mathcal I}_k = {\mathcal J}_k + {\mathcal O}\left( {{\gamma ^2}} \right).
\end{equation}
In the rest of the work, we only compute ${\mathcal J}_k$. 
We have
\begin{equation} \label{eq: U_k vs. u^(0)}
{U_k}\left( {0,t} \right) = u_k^{\left( 0 \right)}\left( {0,t} \right).
\end{equation}
Now, we substitute (\ref {eq: U_k vs. u^(0)}) and (\ref {eq: first-order perturbation}) into (\ref {eq: J_k xcorr}), to obtain
\begin{equation} \label{eq: J_k xcorr developed}
{\mathcal J}_k = \left\langle {{{\hat {\mathcal D}}_L}\left[ {u_k^{(0)}\left( {0,t} \right)} \right],u^{(0)}\left( {L,t} \right) + \gamma u^{(1)}\left( {L,t} \right)} \right\rangle.
\end{equation}
Now, we write ${\mathcal J}_k$ as the sum of the signal, nonlinear signal-signal distortions, NSNI distortions and ASE,
\begin{equation} \label{eq: J_k decomposition}
{\mathcal J}_k = {a_k} + {b_k} + {c_k} + {n_k},
\end{equation}
where, $a_k$ is the $k$'th symbol of the COI (cf. (\ref {eq: E(0,t) def.})), $b_k$ is the nonlinear signal-signal distortion on the $k$'th symbol of the COI, $c_k$ is the nonlinear signal-noise distortion term on the $k$'th symbol of COI, and $n_k$ is the ASE noise added to the $k$'th symbol of COI. Note that, given Nyquist pulseshaping and normalized matched filtering, the $k$'th received symbol of COI is equal to the $k$'th transmitted symbol of the COI, \emph{i.e.}, $a_k$. Based on the central limit theorem, $b_k$, $c_k$, and $n_k$ are zero-mean Gaussian random variables, \emph{i.e.}, $b_k\sim\mathcal{N}(0,\sigma_{SS}^2)$, $c_k\sim\mathcal{N}(0,\sigma_{NS}^2)$, $n_k\sim\mathcal{N}(0,\sigma_{ASE}^2)$, where, $\sigma_{SS}^2$, $\sigma_{NS}^2$, and, $\sigma_{ASE}^2$, stand for signal-signal, noise-signal, and ASE noise variance respectively. Note that throughout this work, in order to simplify the analysis, we neglect noise-noise interactions in computing $c_k$, although including those terms is straightforward (cf. also (\ref{eq: c_k detailed})).
\subsection{ASE noise variance}
In order to compute the variance of the sampled ASE noise, $n_k$, we can substitute (\ref {eq: zeorth-order complete solution detailed}) into (\ref{eq: J_k xcorr developed}) and use the DEF. We have
\begin{equation} \label{eq: n_k xcorr}
{n_k} = \left\langle {u_k^{\left( 0 \right)}\left( {0,t} \right),{u_{ASE}}\left( {L,t} \right)} \right\rangle.
\end{equation}
Using (\ref{eq: relation of kappa with xi})-(\ref{eq: ASE autocorrelation}) together with (\ref{eq: n_k xcorr}) we obtain
\begin{equation} \label{eq: n_k variance}
\sigma_{ASE}^2=\frac{1}{2\pi}\frac{\hbar\omega_0}{\mathcal E}\int_{ - \infty }^{ + \infty } {{{d\omega|{{{\tilde u}^{(0)}}(0,\omega)}|}^2}\xi(L;\omega)}.
\end{equation}
Now we assume that the ASE spectrum is flat over signal bandwidth, \emph{i.e.}, $\xi(L;\omega)\approx\xi(L;0)$. We have 
\begin{equation} \label {eq: ASE variance final}
\sigma_{ASE}^2(P) = \frac{\sigma_{\text{qn}}^2}{P}\sum\limits_{n = 1}^{N_s}{g_n e^{\delta_n}n_{\text{sp},n}(0)},
\end{equation}
where, $\sigma_{\text{qn}}^2$ is the quantum noise variance over the COI signal bandwidth, \emph{i.e.}, 
\begin{equation} \label{eq: sigma_0^2}
\sigma_{\text{qn}}^2 = \frac{\hbar\omega_0}{T}.
\end{equation}
\subsection{Nonlinear signal-signal distortions}
We can substitute (\ref{eq: first-order solution}) into (\ref{eq: J_k xcorr developed}), and apply the DEF, in order to derive the following expression for the total nonlinear distortion of the sampled photocurrent ${\mathcal J}_k$, \emph{i.e.}, $b_k+c_k$. We have
\begin{multline} \label{eq: b_k + c_k xcorr simplified}
{b_k} + {c_k} = \\
i\gamma {\mathcal E}\int_0^L {dzf\left( z \right){{\left| {{u^{\left( 0 \right)}}\left( {z,t} \right)} \right|}^2}{u^{\left( 0 \right)}}\left( {z,t} \right)u{{_k^{\left( 0 \right)}}^*}\left( {z,t} \right)}. 
\end{multline}
Equation (\ref {eq: b_k + c_k xcorr simplified}) provides an integral representation of the total nonlinear distortions up to first-order in $\gamma$. There exists an abundant literature on computing the variance of signal-signal interactions, $\sigma_{SS}^2$. Our main task in the rest of this paper is to single out contributions to $c_k$ from the right hand side of (\ref {eq: b_k + c_k xcorr simplified}), in order to compute the variance of the NSNI distortions, $\sigma_{NS}^2$; however, for the sake of completeness we first consider signal-signal distortions.

Now, we substitute the zeroth-order solution, (\ref{eq: zeorth-order complete solution detailed with KL}) into (\ref {eq: b_k + c_k xcorr simplified}). After Collecting all signal-signal product terms we find the following expression for the total nonlinear signal-signal distortions\footnote{Throughout this subsection, we freely use the notational convention, introduced in sec. \ref{sec: Notations and basic definitions} on handling the sub-indices of the waveforms: if the second sub-index indicating the channel index is equal to zero, it can be optionally dropped: $x_{k,0}(z,t)=x_k(z,t)$.}
\begin{multline} \label{eq: b_k, detailed expression}
{b_k} = \sum\limits_{m,n,p} {{a_{m + k}}{a_{n + k}}a_{p + k}^*X_{m,n,p}^{\left( {0,0} \right)}}  + \\
2\sum\limits_s {\sum\limits_{m,n,p} {{a_{m + k}}{a_{n + k,s}}a_{p + k,s}^*X_{m,n,p}^{\left( {0,s} \right)}} }  + \\
\sum\limits_{\mathop {s,s'}\limits_{s \ne s'} } {\sum\limits_{m,n,p} {{a_{m + k,s}}{a_{n + k,s'}}a_{p + k,s + s'}^*X_{m,n,p}^{\left( {s,s'} \right)}} }. 
\end{multline}
In (\ref {eq: b_k, detailed expression}) the symbol $\sum\limits_{m,n,p}$ is a short-hand notation for $\sum\limits_{m =  - \infty }^{ + \infty } {\sum\limits_{n =  - \infty }^{ + \infty } {\sum\limits_{p =  - \infty }^{ + \infty }} }$. By $\sum\limits_s$ we mean $\sum\limits_{s,s \ne 0}$. The same convention holds for the double sum on $s$ and $s'$. The perturbative coefficients in the expansion (\ref {eq: b_k, detailed expression}) are
\begin{equation} \label{eq: X_mnp definition}
X_{m,n,p}^{\left( {s,s'} \right)} = i\gamma {\mathcal E}\int_0^L {dzK_{m,n,p}^{\left( {s,s'} \right)}\left( z \right)},
\end{equation}
where, the integral kernels are explicitly written as
\begin{multline} \label{eq: K_mnp definition}
K_{m,n,p}^{(s,s')}(z) = f(z)e^{i[\phi_s(z)+\phi_{s'}(z)-\phi_{s + s'}(z)]}\langle u_0^{(0)}(z,t),\\
u_{m,s}^{(0)}(z,t-t_s)u_{n,s'}^{(0)}(z,t-t_{s'})u_p^{(0)*}(z,t-t_{s + s'})\rangle. 
\end{multline}
The first, second and third sums in (\ref {eq: b_k, detailed expression}) correspond to intra-channel, degenerate inter-channel and non-degenerate inter-channel signal-signal FWM terms. At high symbol-rates (say 28 GBaud and beyond) the contribution of the non-degenerate FWM (NDFWM) terms to the total variance of nonlinear distortions is smaller than the other two terms in (\ref{eq: b_k, detailed expression}); however, at low symbol-rates, \emph{e.g.}, in the case of subcarrier multiplexing, \cite{Guiomar2017 : OptXpress2017}, the NDFWM becomes important. In this paper we keep the NDFWM terms for the sake of completeness. We will examine the impact of NDFWM on the performance of both uncompensated and compensated system performance in the next section. 
The derivation of the variance of the signal-signal terms is discussed in \cite{Dar2013 : OptXpress2013, Dar2014 : OptXpress2014, Dar2015 : JLT2015, Dar2016 : JLT2016}. Here for the reference we write only the end result, which is
\begin{multline} \label{eq: sig^2_SS variance}
\sigma_{SS}^2(P)=\\
\gamma^2P^2\{2\mathcal{X}_1 + (\frac{\mu_4}{\mu_2^2}-2) [\mathcal{X}_2 + 4\mathcal{X}_3 + 4\mathcal{X}_4] + \\
(\frac{\mu_6}{\mu_2^3}-9\frac{\mu_4}{\mu_2^2} + 12)\mathcal{X}_5 + 4\sum\limits_s[\mathcal{X}_{1,s}+(\frac{\mu_4}{\mu_2^2}-2)\mathcal{X}_{3,s}] + \\
\sum\limits_s \sum\limits_{s'} \mathcal{X}_{1,s,s'}\}
\end{multline}
The various $\mathcal{X}$-coefficients appearing in (\ref {eq: sig^2_SS variance}) are calculated in the Appendix. Note that $\mathcal{X}_2$ and $\mathcal{X}_4$ turn out to be respectively one and two orders of magnitude smaller than $\mathcal{X}_3$, and can be neglected in computing the signal-signal variance without impacting its numerical value. 
\subsection{Nonlinear signal-noise distortions}
When (\ref {eq: zeorth-order complete solution detailed with KL}) is substituted into (\ref {eq: b_k + c_k xcorr simplified}), many product terms are resulted. All the terms that are expressed as the product of three symbols contribute to the signal-signal distortions, and are taken into account in (\ref{eq: b_k, detailed expression}). All the other terms, which are the products of either one or two symbols with either two or one ASE KL coefficients model the parametric amplification of noise by signal, and contribute to the NSNI. These product terms are first-order, or second-order in noise respectively. There are also product terms among three ASE noise coefficients, which contribute to the nonlinear noise-noise interactions. In the following we are considering only the leading NSNI terms that are first-order in ASE coefficients. We have
\begin{multline} \label{eq: c_k detailed}
{{c}_k} = i\gamma {\mathcal E}\int_0^L {dz}\sum\limits_{m,n,p}\{\\
{[{{a_{m+k}}{a_{n+k}}w_{p+k}^*\left( z \right) + 2{a_{m+k}}a_{p+k}^*{w_{n+k}}(z)}]K_{m,n,p}^{(0,0)}(z)} \\
 + 2\sum\limits_s [{a_{m+k}}{a_{n+k,s}}w_{p+k,s}^*(z) + {a_{m+k}}a_{p+k,s}^*{w_{n+k,s}}(z)\\
 + {a_{n+k,s}}a_{p+k,s}^*{w_{m+k}}(z)]K_{m,n,p}^{(0,s)}(z)\\
+ \sum\limits_{s,s'}[{a_{m+k,s}}{a_{n+k,s'}}w_{p+k,s+s'}^*(z)\\ 
+ {a_{m+k,s}}a_{p+k,s+s'}^*{w_{n+k,s'}}(z)\\
+ {a_{n+k,s'}}a_{p+k,s+s'}^*{w_{m+k,s}}(z)]K_{m,n,p}^{(s,s')}(z)\}.
\end{multline}
In writing (\ref {eq: c_k detailed}) we have used the fact that, (cf. (\ref{eq: K_mnp definition})),
\begin{equation}\label{eq: K_mnp^(0, 0) symmetry}
 K_{m,n,p}^{\left( {0,0} \right)}\left( z \right) = K_{n,m,p}^{\left( {0,0} \right)}\left( z \right).
\end{equation}
Now we use the detailed expression for $c_k$, as per (\ref{eq: c_k detailed}), to compute $\sigma_{NS}^2=\langle|c_k|^2\rangle=\langle|c_0|^2\rangle$. After some straightforward algebra we derive
\begin{multline} \label{eq: variance c_k detailed}
\sigma_{NS}^2 = \gamma^2\mathcal E^2\int_0^L {dz} \int_0^L {dz'}\sum\limits_{m,n,p}\sum\limits_{m',n',p'}\{\\
{\mathcal W}_{m',n',p'}^{m,n,p}(z,z')K_{m,n,p}^{(0,0)}(z)K{_{m',n',p'}^{(0,0)*}}(z')\\
+ \sum\limits_s{}{\mathcal W}_{m',n',p',s}^{m,n,p}(z,z')K_{m,n,p}^{(0,s)}(z)K{_{m',n',p'}^{(0,s)*}}(z')\\
+ \sum\limits_{s,s'}{}{\mathcal W}_{m',n',p',s,s'}^{m,n,p}(z,z')K_{m,n,p}^{(s,s')}(z)K{_{m',n',p'}^{(s,s')*}}(z')\},
\end{multline}
where, the nonlinear kernels are multiplied with the weight functions, which are space-dependent ensemble averages over WDM symbols and ASE noise coefficients. The weight function for the intra-channel FWM of the COI is 
\begin{multline} \label{eq: W def.}
{\mathcal W}_{m',n',p'}^{m,n,p}\left( {z,z'} \right) = 
4\left\langle {{a_m}a_p^*{a_{m'}}^*a_{p'}^{}} \right\rangle \left\langle {{w_n}\left( z \right)w_{n'}^*\left( {z'} \right)} \right\rangle \\
 + \left\langle {{a_m}{a_{m'}}^*{a_n}{a_{n'}}^*} \right\rangle \left\langle {w_p^*\left( z \right)w_{p'}^{}\left( {z'} \right)} \right\rangle, 
\end{multline}
For the degenrate FWM between COI and the $s^{\text{th}}$  channel the weight function is
\begin{multline} \label{eq: W_s def.}
{\mathcal W}_{m',n',p',s}^{m,n,p}\left( {z,z'} \right) = \\
4\left\langle {{a_m}{a_{m'}}^*} \right\rangle \left\langle {a_{p,s}^*a_{p',s}^{}} \right\rangle \left\langle {{w_{n,s}}\left( z \right)w_{n',s}^*\left( {z'} \right)} \right\rangle \\
 + 4\left\langle {{a_m}{a_{m'}}^*} \right\rangle \left\langle {{a_{n,s}}{a_{n',s}}^*} \right\rangle \left\langle {w_{p,s}^*\left( z \right)w_{p',s}^{}\left( {z'} \right)} \right\rangle \\
 + 4\left\langle {{a_{n,s}}{a_{n',s}}^*}{a_{p,s}^*a_{p',s}^{}} \right\rangle \left\langle {{w_m}\left( z \right)w_{m'}^*\left( {z'} \right)} \right\rangle \\, 
\end{multline}
and finally, for the NDFWM among COI, $s^{\text{th}}$  and $s'^{\text{th}}$  channel the weight function is
\begin{multline} \label{eq: W_s,s' def.}
{\mathcal W}_{m',n',p',s,s'}^{m,n,p}\left( {z,z'} \right) = \\
\left\langle {{a_{m,s}}{a_{m',s}}^*} \right\rangle \left\langle {a_{p,s+s'}^*a_{p',s+s'}^{}} \right\rangle \left\langle {{w_{n,s'}}\left( z \right)w_{n',s'}^*\left( {z'} \right)} \right\rangle \\
 +\left\langle {{a_{m,s}}{a_{m',s}}^*} \right\rangle \left\langle {{a_{n,s'}}{a_{n',s'}}^*} \right\rangle \left\langle {w_{p,s+s'}^*\left( z \right)w_{p',s+s'}^{}\left( {z'} \right)} \right\rangle \\
 +\left\langle {{a_{n,s'}}{a_{n',s'}}^*}\rangle\langle{a_{p,s+s'}^*a_{p',s+s'}^{}} \right\rangle \left\langle {{w_{m,s}}\left( z \right)w_{m',s}^*\left( {z'} \right)} \right\rangle \\.
\end{multline}
In order to calculate these weights the ensemble averages should be computed. In order to do so, we assume that the WDM symbols are zero-mean random variables, and that symbols at different symbol time intervals are i.i.d. The $n^{\text{th}}$ moment of the symbols is denoted by $\mu_n$. We assume that all WDM channels are modulated with the same modulation format. The $n^{\text{th}}$ moment of the constellation is
\begin{equation} \label{eq: mu_4}
{\mu _n} = \left\langle {{{\left| {{a_m}} \right|}^n}} \right\rangle  = \left\langle {{{\left| {{a_{m,s}}} \right|}^n}} \right\rangle.
\end{equation}
In this work, we assume that $\mu_1=0$. Given the symbols are i.i.d., the four-times moment of the COI symbol is
\begin{multline} \label{eq: four-fold symbols xcorr.}
\left\langle {{a_m}a_p^*{a_{m'}}^*a_{p'}^{}} \right\rangle  = \\
\mu _2^2{\delta _{mm'}}{\delta _{pp'}} + \mu _2^2{\delta _{mp}}{\delta _{m'p'}} + \left( {{\mu _4} - 2\mu _2^2} \right){\delta _{mm'}}{\delta _{pp'}}{\delta _{mp}},
\end{multline}
and similarly, the four-time moment of the $s^{\text{th}}$ channel is 
\begin{multline} \label{eq: four-fold symbols xcorr. ch. s}
\left\langle {{a_{n,s}}a_{p,s}^*{a_{n',s}}^*a_{p',s}^{}} \right\rangle  = \\
\mu _2^2{\delta _{nn'}}{\delta _{pp'}} + \mu _2^2{\delta _{np}}{\delta _{n'p'}} + \left( {{\mu _4} - 2\mu _2^2} \right){\delta _{nn'}}{\delta _{pp'}}{\delta _{np}}.
\end{multline}
The variance of the KL expansion coefficients of the ASE stochastic process was computed in (\ref {eq: w_k autocorr aprrox}) and (\ref {eq: w_k,s autocorr approx}). Here we write again the explicit expressions for two-times ensembale average of the ASE KL expansion coefficients with arbitrary indices. For the ASE coefficients added to the COI we have
\begin{equation} \label{eq: w variance}
\left\langle {{w_m}\left( z \right)w_{m'}^*\left( {z'} \right)} \right\rangle  = {\delta _{mm'}}\kappa \left( {z,z';0} \right),
\end{equation}
and, for the ASE coefficients added to the $s^{\text{th}}$ channel we obtain
\begin{equation} \label{eq: w_s variance}
\left\langle {{w_{m,s}}\left( z \right)w_{m',s}^*\left( {z'} \right)} \right\rangle  = {\delta _{mm'}}\kappa \left( {z,z';{\Omega _s}} \right).
\end{equation}
Now we substitute (\ref{eq: four-fold symbols xcorr.})-(\ref {eq: w_s variance}) into the expressions for the kernel wieghtrs, \emph{i.e.}, equations (\ref {eq: W def.}), (\ref {eq: W_s def.}), and (\ref {eq: W_s,s' def.}). The resulting expressions for the kernel weights are then substituted into (\ref {eq: variance c_k detailed}). After simplifications we finally find the following expression for the total variance of the NSNI distortions
\begin{multline} \label{eq: c_k variance in terms of Chi's}
\sigma_{NS}^2(P) = \gamma ^2\sigma_{\text{qn}}^2P\{6\chi _1+(\frac{\mu _4}{\mu _2^2}-2)(\chi_2+4\chi_3)\\
 + 4\sum\limits_s(\chi_{1,s}+2\chi'_{1,s})+(\frac{\mu _4}{\mu _2^2}-2)\sum\limits_s \chi_{3,s} \\
 + \sum\limits_s\sum\limits_{s'}(\chi'_{1,s}+\chi'_{1,s'}+\chi'_{1,s+s'})\}.
\end{multline}
The various $\chi$-coefficients in (\ref {eq: c_k variance in terms of Chi's}) are discussed in detail in the appendix, where, in the first step they are represented as multi-dimensional integrals, and then those integrals are cast in a form that can be efficiently computed by Monte Carlo integration. 
\subsection{Extension to dual polarization}
Up to here, we assumed that the WDM field in the fiber is perfectly polarized, and that its propagation is governed by the NLS. In this subsection we show how the expressions derived previously for the the variance of the ASE, signal-signal, and signal-noise distortions can be extended to the case of dual-polarization signals. We assume that signal propagation in the dual-polarization case is modeled by the Manakov equation, which is
\begin{equation} \label{eq: Manakov equation}
\partial_z\mathbf{E}=\frac{{g\left( z \right) - \alpha \left( z \right)}}{2}{\mathbf{E}} - i\frac{{{\beta _2}\left( z \right)}}{2}\partial _t^2{\mathbf{E}} + i\frac{8}{9}\gamma {{\mathbf{E}}^\dag }{\mathbf{EE}} + {\mathbf{n}}\left( {z,t} \right),
\end{equation}
where, $\mathbf{E}=[E_x(z,t),E_y(z,t)]^T$ and $\mathbf{n}(z,t)=[n_x(z,t),n_y(z,t)]^T$. The subscripts $x$ and $y$ refer to the two transverse components of the electromagnetic field, the superscript $T$ stands for matrix transpose operation and the superscript $\dag$ stands for Hermitian conjugation operation, \emph{i.e.}, complex conjugation followed by matrix transpose.
In order to extend the results of the previous sections to the case of a dual-polarization electromagnetic field obeying the Manakov equation, first, we have to change $\gamma$ to $\frac{8}{9}\gamma$, and $P$, to $\frac{1}{2}P$ everywhere in the derivations of the formulae for the single-polarization case, next, we have to take into account additional contributions from cross-polarization terms, \emph{i.e.}, $|E_y|^2E_x$ and $|E_x|^2E_y$ in Manakov equation to the variance of nonlinear distortions.
Let's denote the dual-polarization variance of the ASE, which, in this paper is normalized to the signal power, by $\sigma _{ASE, DP}^2(P)$. Based on (\ref{eq: ASE variance final}) we have
\begin{equation} \label{eq: sig2_NS_DP}
\sigma _{ASE,DP}^2(P)=2\sigma _{ASE}^2(P).
\end{equation}  
The variance of nonlinear signal-signal distortions assuming dual polarizations is denoted by $\sigma _{SS,DP}^2(P)$. We have 
\begin{equation} \label{eq: sig2_SS_DP}
\sigma _{SS,DP}^2\left( P \right) = \frac{{16}}{{81}}\left[ {\sigma _{SS}^2\left( P \right) + \sigma _{SS,XP}^2\left( P \right)} \right],
\end{equation}
where, $\sigma_{SS,XP}^2$ is the variance of the signal-signal cross-polarization distortions, which is
\begin{multline} \label{eq: sig2_SS_XP}
\sigma _{SS,XP}^2\left( P \right) = {\gamma ^2}{P^2}\left[ {{\mathcal{X}_1} + \left( {\frac{{{\mu _4}}}{{\mu _2^2}} - 2} \right){\mathcal{X}_3} + \sum\limits_s {{\mathcal{X}_{1,s}}}  + } \right.\\
\left. {\left( {\frac{{{\mu _4}}}{{\mu _2^2}} - 2} \right)\sum\limits_s {{\mathcal{X}_{3,s}}}  + \sum\limits_{s,} {\sum\limits_{s'} {{\mathcal{X}_{1,s,s'}}} } } \right].
\end{multline}
Similarly, let's denote the variance of the total NSNI distortions assuming dual polarizations by $\sigma _{NS,DP}^2(P)$. We have   
\begin{equation} \label{eq: sig2_NS_DP}
\sigma _{NS,DP}^2\left( P \right) = \frac{{32}}{{81}}\left[ {\sigma _{NS}^2\left( P \right) + \sigma _{NS,XP}^2\left( P \right)} \right],
\end{equation}
where, $\sigma _{SN,XP}^2$ is the variance of the signal-noise cross-polarization distortions
\begin{multline} \label{eq: sig2_NS_XP}
\sigma _{NS,XP}^2\left( P \right) = {\gamma ^2}\sigma _{{\text{qn}}}^2P\left[ {3{\chi _1} + \left( {\frac{{{\mu _4}}}{{\mu _2^2}} - 2} \right){\chi _3} + } \right.\\
\left. {3\sum\limits_s {{\chi _{1,s}}}  + \left( {\frac{{{\mu _4}}}{{\mu _2^2}} - 2} \right)\sum\limits_s {{\chi _{3,s}}}  + 3\sum\limits_{s,} {\sum\limits_{s'} {{\chi _{1,s,s'}}} } } \right].
\end{multline}

Finally, putting it all together, the variance terms computed so far have to be substituted in the expressions for the uncompensated, \emph{i.e.}, when no nonlinear compensation equalization is applied, as well as the full-field compensated signal-to-noise ratios (SNR). The uncompensated SNR, denoted by $SNR_U$ is
\begin{equation}\label{eq: SNR_U definition}
SN{R_U}\left( P \right) = \frac{1}{{\sigma _{ASE,DP}^2\left( P \right) + \sigma _{SS,DP}^2\left( P \right) + \sigma _{NS,DP}^2\left( P \right)}},
\end{equation}
and the compensated SNR, denoted by $SNR_C$ is
\begin{equation}\label{eq: SNR_C definition}
SN{R_C}\left( P \right) = \frac{1}{{\sigma _{ASE,DP}^2\left( P \right) + \sigma _{NS,DP}^2\left( P \right)}}.
\end{equation}

\section{Numerical results}
In this section we present numerical results to validate the theory developed in the previous sections. We performed standard PDM WDM numerical simulations using split-step Fourier method (SSFM), and computed $SNR_U(P)$, and $SNR_C(P)$ as a function of channel launch power $P$, and compared the numerical results with the analytical curves, \emph{i.e.}, Eqs. (\ref{eq: SNR_U definition}) and (\ref{eq: SNR_C definition}). Two link configurations are considered. The first configuration, which is 40 spans of 120 km NZDSF fiber and 49 GBd PDM-QPSK, with 50 GHz spacing, is similar to \cite{Poggiolini : ECOC2014}, where the impact of in-line ASE is considerable even without nonlinear compensation. The second configuration is a typical terrestrial scenario, where the link is composed of 20 spans of 100 km SMF fiber, and 49 GBd PDM-16QAM with 50 GHz spacing is assumed. Tab.\ref{tab: simulation params.} contains all the relevant system parameters for these two simulations scenarios. For each configuration we considered WDM transmission with 1, 5, and 15 channels. For the second configuration we also examined WDM transmission with 89 channels.       
\begin{table}[!t]
\renewcommand{\arraystretch}{1.0}
\caption{Simulation parameters}
\label{tab: simulation params.}
\centering
\begin{tabular}{|c||c||c|}
\hline
& Config. 1 & Config. 2\\
\hline
Fiber type & NZDSF & SMF\\
\hline
Span length & 120 km & 100 km\\
\hline
No. spans & 40 & 20\\
\hline
No. WDM channels & 1, 5, 15 & 1, 5, 15, 89\\
\hline
Modulation format & PDM-QPSK & PDM-16QAM\\
\hline
Symbol-rate & 49 GBd & 49 GBd\\
\hline
Channel spacing & 50 GHz & 50 GHz\\
\hline
Attenuation coeff. $\alpha$ & 0.22 [dB/km] & 0.20 [dB/km] \\
\hline
Dispersion coeff. $D$ & 3.8 [ps/nm/km] & 16.5 [ps/nm/km]\\
\hline
Fiber slope & 0 & 0\\
\hline
Effective area $A_{eff}$ & 70.26 [$\mu m^2$] & 80 [$\mu m^2$] \\
\hline
Central wavelength $\lambda_0$ & 1550 [nm] & 1550 [nm]\\
\hline
EDFA noise figure & 5 dB & 5 dB\\
\hline
pulse shape & RRC, roll-off 0.001 & RRC roll-off 0.001\\
\hline
\end{tabular}
\end{table}  
In all cases full-field DBP was applied at the receiver side. Other configurations of DBP, \emph{e.g.}, pre-compensation, or mixed pre-and post-compensation will be examined in future. Transmitter and receiver are assumed ideal with zero quantization noise. The receiver consisted only of matched filter, constellation rotation, and SNR computation by comparing transmitted and received signals. The pulse-shape was root-raised-cosine (RRC) with roll-off 0.001. In order to calibrate the numerical simulator noiseless transmission followed by full-field DBP was performed at channel power equal to 8 dBm, which was far beyond the nonlinear threshold. With roll-off 0.001 we measured SNRs about 400 dB (instead of infinite) whereas roll-off 0 resulted in an SNR around 40 dB. We therefore decided to set the roll-off to 0.001 in all numerical simulations to guarantee that the residual numerical error is negligible and does not influence the numerical results.   

\begin{figure*}[!t]
\centering
\includegraphics[width=7in]{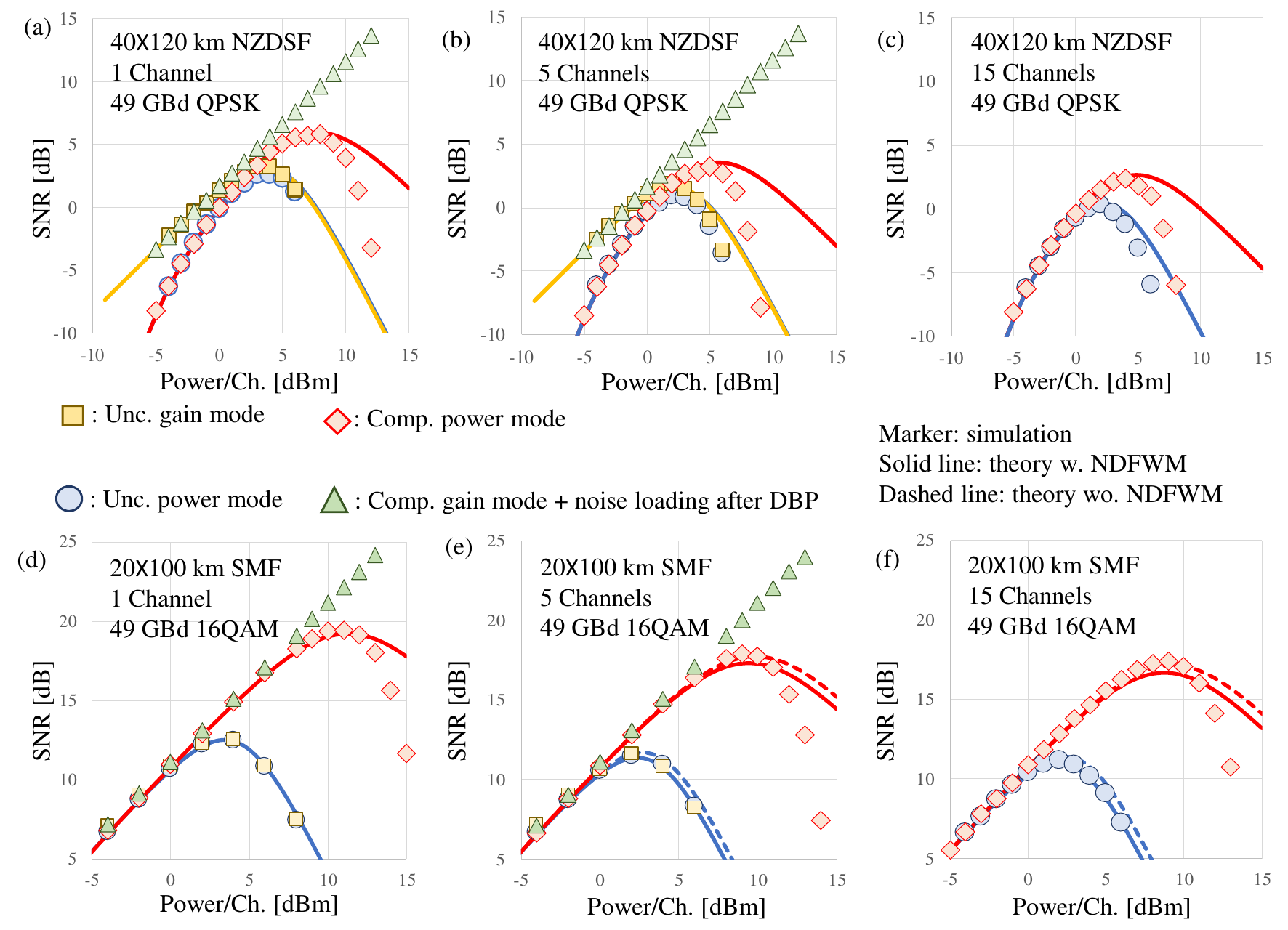}
\hfil
\caption{Comparison between numerical simulations (markers), and theoretical curves (lines), top row: configuration 1, bottom row configuration. 2. For each configuration, PDM WDM transmission with 1, 5, and 15 channels are considered. Unc.: uncompensated (i.e. no full-field DBP), Comp.: compensated (i.e. full-field DBP applied at receiver side). For 1- and 5-channel scenarios uncompensated transmission is performed both in gain mode and in power mode. }
\label{fig: Fig.1 }
\end{figure*}
Figure 1 illustrates the numerical and theoretical SNR vs. per channel lunch power for the two configurations, and for 1, 5, and 15 WDM channels. Fig. 1 top row corresponds to configuration 1, and Fig. 1 bottom row corresponds to configuration 2. Fig. 1a shows the SNR curves of single-channel transmission of configuration 1. The blue circles are numerical results corresponding to the uncompensated (Unc.) transmission where all EDFAs operate at constant output power (power mode), and the yellow squares correspond to the uncompensated transmission case where EDFAs deliver constant signal gain to fully compensate the span loss (gain mode). The solid lines are the theoretical SNR vs. channel power curves as per Eqs. (\ref{eq: SNR_U definition}) and (\ref{eq: SNR_C definition}). Red diamonds are numerical results when full-field DBP is applied at the receiver side to the waveforms in the power mode. We also applied full-field compensation to the waveforms in the gain mode, but the results are not shown here in order to avoid cluttering the figures. For sanity check we also did numerical simulations with noiseless transmission followed by full-field DBP, and then loaded the noise after the DBP. This case was simulated to make sure that in absence of noise the nonlinear channel is perfectly inverted. The results of this case are shown in green triangles, and confirm that in the absence of in-line noise, the full-field zero-forcing DBP nonlinear compensation is perfect. In configuration 1 signal depletion by ASE is not negligible, and that is why the uncompensated curves assuming gain mode and power mode are different specially in the linear regime. Both of these uncompensated numerical curves match very well with the corresponding theoretical curves. As to the compensated case, we observe that the theoretical (solid red) curve, Eq. (\ref{eq: SNR_C definition}), matches the numerical results (red diamonds) up to the optimum power, but then there is a discrepancy between the theoretical and the numerical curves. The same behavior is observed (but not shown in Fig. 1) when compensation is applied to waveforms transmitted the gain mode. The explanation of this divergence is the following: if the nonlinear channel is perfectly inverted, signal-signal interactions are canceled to all orders. We have verified that this is so in noiseless transmission followed by full-field DBP (green triangles), which corresponds to when the compensated SNR would be $SNR_C(P) = \sigma^{-2}_{ASE,DP}$ instead of (\ref{eq: SNR_C definition}); however, if in-line noise is present, the power profile applied in the backpropagation is not exactly the inverted version of the power profile  of the forward propagation part, and this asymmetry in the forward and backward portions of signal propagation results in the presence of the residual signal-signal nonlinear distortions, whereas the theoretical curve based on (\ref{eq: SNR_C definition}) assumes signal-signal nonlinear distortions are perfectly compensated to all orders. Interestingly, our theory well approximates the compensated performance up to the optimum point, and is sufficient to evaluate the fundamental zero-forcing limit of full-field DBP. Figs. 1b and 1c illustrate numerical and theoretical SNR vs. power curves assuming configuration 1, but with 5 and 15 WDM transmitted channels. Figs. 1b and 1c show the SNRs of the third among five, and eight among fifteen channels respectively. similar trends are observed in Figs. 1a, 1b, and 1c. We have thus verified that our theory successfully predicts the performance of the uncompensated transmission in a case where signal depletion by in-line ASE is significant. We also verified that our theory is sufficient to predict system performance assuming full-field DBP up to the optimum power. Figs. 1d, 1e, and 1f illustrate numerical and theoretical SNR vs. power curves in configuration 2. In this scenario signal depletion by ASE is negligible, and uncompensated performance in gain mode and power mode are essentially the same. For configuration 2, we plotted the theoretical curves, both with NDFWM contributions included, solid lines, and with NDFWM terms excluded, dashed lines, in order to assess the importance of NDFWM. the NDFWM terms do not exist in single-channel transmission, Fig. 1d. We observed in Fig. 1e and 1f that excluding NDFWM results in over-estimating the SNR by 0.5 dB and 0.7 dB respectively. We also observe that the red solid curves corresponding to the complete theory with NDFWM included slightly underestimate the numerical curves although the underestimation is about 0.5 dB. This might be partly due to the statistical uncertainty of the estimate numerical SNR, and the limited accuracy of the numerical simulator, and partly due to the inherent approximations in our theory. This issue will be explored further in future research.  

\begin{figure}[!t]
\centering
\includegraphics[width=3.5in]{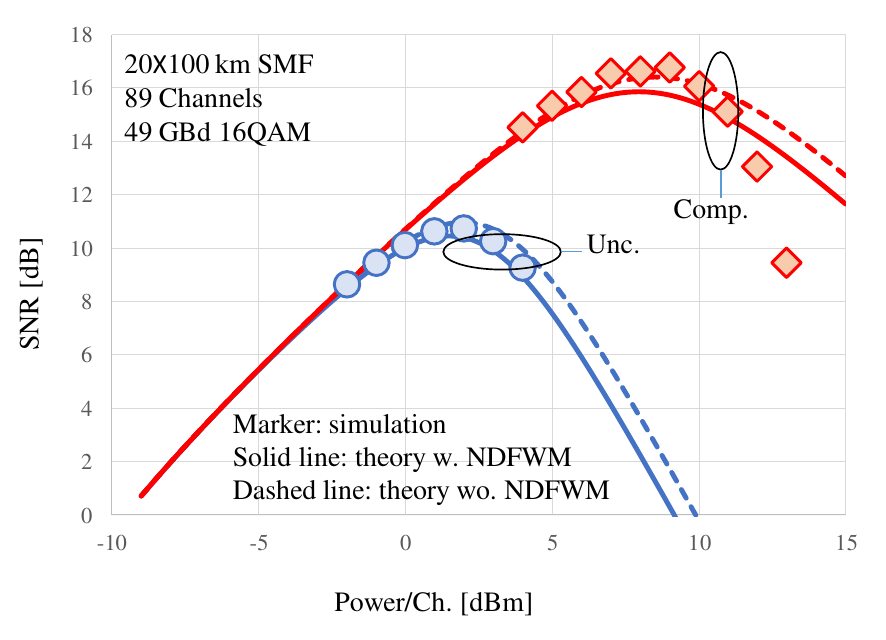}
\hfil
\caption{SNR vs. power of the central channel in 89-channel transmission simulation of configuration 2.}
\label{fig_sim}
\end{figure}

Fig.2 illustrates numerical and theoretical SNR vs. power curves in configuration 2, but when 89 WDM channels are transmitted. The COI is the central, \emph{i.e.}, the forty fifth channel. The massive WDM simulations in the compensated case took two weeks to finish using an NVIDIA Tesla K40 GPU card.  

\begin{figure}[!t]
\centering
\includegraphics[width=3.5in]{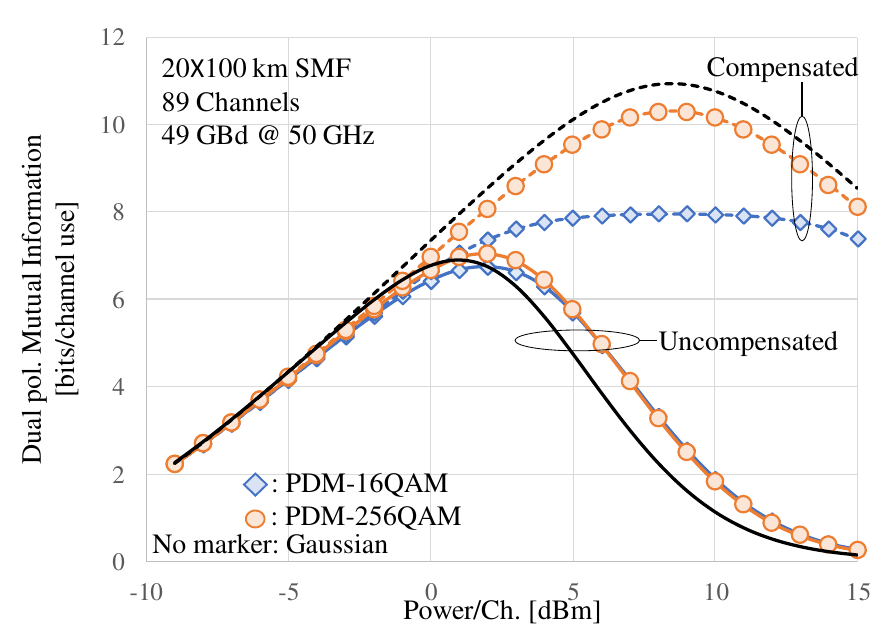}
\hfil
\caption{Dual polarization mutual information of 16QAM, 256QAM and Gaussian constellations in configuration two, with and without full-field digital-back propagation post-compensation. }
\label{fig_sim}
\end{figure}

Fig.3 illustrates the mutual information (MI) vs. channel power in configuration 2 with and without full-field nonlinear compensation. We examined 16QAM, 256QAM and constellation, and for the reference, also computed the MI of a Gaussian source. In the uncompensated case the optimum MI of the Gaussian source is less than that of 16QAM and 256 QAM, due to its higher fourth and sixth moments. In the fully-compensated case 16QAM optimum MI saturates to its dual polarization maximum constrained capacity, \emph{i.e.}, 8 bits/channel use, 256QAM optimum MI is 10.3 bits/channel use, and Gaussian source optimum MI is 10.9 bits/channel use.
 
\section{Conclusions}
In this work we presented a rigorous derivation of a general theory of nonlinear signal-noise interactions in WDM fiber-optic coherent transmission systems. This theory is based on the regular perturbation approximation of the nonlinear Schr\"odinger equation, and is exact up to the first-order. The theory is general in that, it is valid for dispersion-managed/un-managed systems, all cross-channel nonlinear FWM terms, and the impact of modulation format are taken into account. Heterogeneous spans with both erbium-doped fiber and Raman amplification are allowed, and chromatic dispersion to all-orders can be included. This theory was applied to compute the total variance of the signal distortions at the receiver. Ideal multi-channel digital backpropagation was optionally included in the receiver side. The variance of the signal distortion was expressed as a sum of various signal-signal and signal-noise contributions. First, integral representations were derived for these contributions, and then these representations were further manipulated to obtain equivalent forms, which could be efficiently computed by Standard Monte Carlo integration. The validity of the developed theory was examined by comparing the theoretical signal-to-noise ratio vs. launched power curves with the corresponding numerical curves from WDM transmission simulations using split-step Fourier method. Two link configurations and four channel counts, (1, 5, 15, and 89) were examined. In all the scenarios the theory could predict the optimum uncompensated and compensated performance up to 0.5 dB of discrepancy. 

\appendix
In this appendix the $\mathcal{X}$-coefficients appearing in the formula for the variance of nonlinear signal-signals distortions in (\ref{eq: sig^2_SS variance}), and the $\chi$-coefficients appearing in the formula for the variance of NSNI in (\ref {eq: c_k variance in terms of Chi's}) are computed. In \ref{app. subsec. A} we derive integral representations for $\mathcal{X}$ and $\chi$-coefficients. In \ref{app. subsec. B} we simplify these integral representations. In \ref{app. subsec. C}, the simplified integral representations are further manipulated to do efficient numerical computations.      
\subsection{Integral representation of $\mathcal{X}$- and $\chi$-coefficients} \label{app. subsec. A}
Consider the following spectral-domain integral representation of the kernel integrals in (\ref{eq: K_mnp definition})
\begin{equation} \label{eq: K^(s,s')_mnp frequency domain}
K_{m,n,p}^{\left( {s,s'} \right)}\left( z \right) = \int_{{\rm I\!R}^3} {\frac{{{d^3}\omega }}{{{{\left( {2\pi } \right)}^3}}}} {H_{{\vec \omega },s,s'}}\left( z \right){e^{i\left( {m{\omega _1} - p{\omega _2} + n{\omega _3}} \right)T}}.
\end{equation}
Throughout this work the symbol $d^n\omega$ stands for $d\omega_1d\omega_2\cdots d\omega_n$ for any positive integer $n$, and the following notational convention is used 
\begin{equation} \label{eq: omega vect. def.}
\vec \omega  = \left( {{\omega _1},{\omega _2},{\omega _3}} \right).
\end{equation}
The integrand function in (\ref{eq: K^(s,s')_mnp frequency domain}) is 
\begin{equation} \label{eq: H_s,s' def.}
H_{\vec\omega,s,s'}(z)=\Pi_{\vec\omega,s,s'}f(z)e^{i\beta _2(z)(\omega_2-\omega_3)(\omega_2-\omega_1)},
\end{equation}
where, the function $\Pi_{\vec\omega,s,s'}$ is defined to be
\begin{multline} \label{eq: Pi_omega def.}
\Pi_{\vec\omega,s,s'} = \tilde u_0^{(0)}(0,\omega_1-\Omega_s)\tilde u{_0^{(0)*}}(0,\omega_2-\Omega_{s+s'})\times \\
\tilde u_0^{(0)}(0,\omega_3-\Omega_{s'})\tilde u{_0^{(0)*}}\left( {0,{\omega_1} - {\omega_2} + {\omega_3}} \right).
\end{multline}
Note that the following notional simplifications are used in the following: $H_{\vec\omega,s}=H_{\vec\omega,0,s}$, $\Pi_{\vec\omega,s}=\Pi_{\vec\omega,0,s}$, $H_{\vec\omega}=H_{\vec\omega,0,0}$, and $\Pi_{\vec\omega}=\Pi_{\vec\omega,0,0}$. 
For the $\mathcal{X}$-coefficients contributing to the variance of nonlinear signal-signal distortions, the following expressions are derived, \cite{Dar2013 : OptXpress2013, Dar2014 : OptXpress2014, Dar2015 : JLT2015, Dar2016 : JLT2016}, 
\begin{equation} \label{eq: X_1 def. basic}
\mathcal{X}_1=\sum\limits_{m,n,p} {{{\left| {X_{m,n,p}^{\left( {0,0} \right)}} \right|}^2}},
\end{equation}
\begin{equation} \label{eq: X_2 def. basic}
\mathcal{X}_2=\sum\limits_{m,n} {{{\left| {X_{m,m,n}^{\left( {0,0} \right)}} \right|}^2}},
\end{equation}
\begin{equation} \label{eq: X_3 def. basic}
\mathcal{X}_3=\sum\limits_{m,n} {{{\left| {X_{m,n,n}^{\left( {0,0} \right)}} \right|}^2}},
\end{equation}
\begin{equation} \label{eq: X_4 def. basic}
\mathcal{X}_4=\operatorname{Re} \left\{ {\sum\limits_{m,n} {X_{m,n,n}^{\left( {0,0} \right)}X{{_{m,m,m}^{\left( {0,0} \right)}}^*}} } \right\},
\end{equation}
\begin{equation} \label{eq: X_5 def. basic}
\mathcal{X}_5=\sum\limits_m {{{\left| {X_{m,m,m}^{\left( {0,0} \right)}} \right|}^2}},
\end{equation}
\begin{equation} \label{eq: X_1,s def. basic}
\mathcal{X}_{1,s}=\sum\limits_{m,n,p} {{{\left| {X_{m,n,p}^{\left( {0,s} \right)}} \right|}^2}},
\end{equation}
\begin{equation} \label{eq: X_3,s def. basic}
\mathcal{X}_{3,s}=\sum\limits_{m,n} {{{\left| {X_{m,n,n}^{\left( {0,s} \right)}} \right|}^2}},
\end{equation}
\begin{equation} \label{eq: X_1,s,s' def. basic}
\mathcal{X}_{1,s,s'}=\sum\limits_{m,n,p} {{{\left| {X_{m,n,p}^{\left( {s,s'} \right)}} \right|}^2}}.
\end{equation}
We can replace (\ref{eq: K^(s,s')_mnp frequency domain}) into (\ref{eq: X_1 def. basic}) to obtain an integral representation for $\mathcal{X}_1$, which is written as follows
\begin{equation} \label{eq: X_1 integral representation}
\mathcal{X}_1=T^2\int_0^L {dz} \int_0^L {dz'} \sum\limits_{m,n,p} {K_{m,n,p}^{\left( {0,0} \right)}\left( z \right)K{{_{m,n,p}^{\left( {0,0} \right)}}^*}\left( {z'} \right)}.
\end{equation}
Similar expressions can be obtained for all other $\mathcal{X}$-coefficients\footnote{as the signal-signal distortions have been extensively studied in the literature we do not present the integral representation of the $\mathcal{X}$-coefficients other than $\mathcal{X}_1$ to save space.}.

In order to simplify the $\chi$-coefficients contributing to the variance of nonlinear signal-noise distortions, we start from (\ref{eq: variance c_k detailed}) and (\ref{eq: c_k variance in terms of Chi's}), and Eqns. (\ref{eq: W def.})-(\ref{eq: w_s variance}). The following simplified integral representations are resulted
\begin{multline} \label{eq: chi_0 def.}
{\chi _0} = T^2\int_0^L {dz} \int_0^L {dz'}\xi(\min(z,z');0) \times \\
\sum\limits_{m,n,p} {} K_{m,n,n}^{\left( {0,0} \right)}\left( z \right)K_{m,p,p}^{(0,0)*}\left( {z'} \right),
\end{multline}
\begin{multline} \label{eq: chi_1 def.}
{\chi _1} = T^2\int_0^L {dz} \int_0^L {dz'} \xi(\min(z,z');0) \times \\
\sum\limits_{m,n,p} K_{m,n,p}^{(0,0)}(z)K{_{m,n,p}^{(0,0)*}}(z'),
\end{multline}
\begin{multline} \label{eq: chi_2 def.}
{\chi _2} = T^2\int_0^L {dz} \int_0^L {dz'}\xi(\min(z,z');0) \times \\
\sum\limits_{m,n} {} K_{m,m,n}^{\left( {0,0} \right)}\left( z \right)K{_{m,m,n}^{(0,0)*}}\left( {z'} \right),
\end{multline}
\begin{multline} \label{eq: chi_3 def.}
{\chi _3} = T^2\int_0^L {dz} \int_0^L {dz'} \xi(\min(z,z');0)\times \\
\sum\limits_{m,n} {} K_{m,n,n}^{\left( {0,0} \right)}\left( z \right)K{_{m,n,n}^{(0,0)*}}\left( {z'} \right),
\end{multline}
\begin{multline} \label{eq: chi_0,s def.}
{\chi _{0,s}} = T^2\int_0^L {dz} \int_0^L {dz'}\xi(\min(z,z');0) \times \\
\sum\limits_{m,n,p} {} K_{m,n,n}^{\left( {0,s} \right)}\left( z \right)K{_{m,p,p}^{(0,s)*}}\left( {z'} \right),
\end{multline}
\begin{multline} \label{eq: chi_1,s def.}
{\chi_{1,s}} = T^2\int_0^L {dz} \int_0^L {dz'} \xi(\min(z,z');0)\times \\
\sum\limits_{m,n,p} {} K_{m,n,p}^{\left( {0,s} \right)}\left( z \right)K{_{m,n,p}^{(0,s)*}}\left( {z'} \right),
\end{multline}
\begin{multline} \label{eq: chi'_1,s def.}
{\chi'_{1,s}} = T^2\int_0^L {dz} \int_0^L {dz'} \xi(\min(z,z');\Omega_s) \times \\
\sum\limits_{m,n,p} {} K_{m,n,p}^{\left( {0,s} \right)}\left( z \right)K{_{m,n,p}^{(0,s)*}}\left( {z'} \right)
\end{multline}
\begin{multline} \label{eq: chi_3,s def.}
{\chi _{3,s}} = T^2\int_0^L {dz} \int_0^L {dz'} \xi(\min(z,z');0)\times \\
\sum\limits_{m,n} {} K_{m,n,n}^{\left( {0,s} \right)}\left( z \right)K{_{m,n,n}^{(0,s)*}}\left( {z'} \right),
\end{multline}
\subsection{Simplifying integral representations}\label{app. subsec. B}
The goal of this subsection is to simplify the integral representation of the $\mathcal{X}$- and $\chi$-coefficients derived in the previous subsection, \emph{i.e.}, Eqs. (\ref{eq: X_1 integral representation})-(\ref{eq: chi_3,s def.}), by carrying out the discrete sums over symbol indices. Following the idea introduced in\cite{Dar2013 : OptXpress2013, Dar2014 : OptXpress2014, Dar2015 : JLT2015, Dar2016 : JLT2016} let's consider the the following identity
\begin{equation} \label{eq: delta train}
\sum\limits_{m =  - \infty }^{ + \infty } {{e^{im\omega T}}}  = \frac{{2\pi }}{T}\sum\limits_{m =  - \infty }^{ + \infty } {\delta \left( {\omega  - \frac{{2\pi m}}{T}} \right)},
\end{equation}
as well as the following orthogonality condition, which holds for Nyquist pulses
\begin{equation} \label{eq: Nyquist orthogonality in freq. domain}
\tilde u_0^{\left( 0 \right)}\left( {0,\omega } \right)\tilde u{_0^{(0)*}}\left( {0,\omega  - \frac{{2\pi n}}{T}} \right) = {\left| {\tilde u_0^{\left( 0 \right)}\left( {0,\omega } \right)} \right|^2}{\delta _{0n}}.
\end{equation}
Using these two basic relations, the discrete summations can be carried out in Eqs. (\ref{eq: X_1 integral representation})-(\ref{eq: chi_3,s def.}). The following simplified integral representations are found for the $\mathcal{X}$-coefficients appearing in the variance of the signal-signal distortions, (\ref{eq: sig^2_SS variance}),
\begin{equation} \label{eq: X_1 def.}
\mathcal{X}_1=\frac{1}{T}\int_{{\rm I\!R}^4} {\frac{{{d^3}\omega }}{{{{\left( {2\pi } \right)}^3}}}\int_0^L {dz{H_{\vec \omega }}\left( z \right)} } \int_0^L {dz'} H_{\vec \omega }^*\left( {z'} \right),
\end{equation}
\begin{equation} \label{eq: X_2 def.}
\mathcal{X}_2=\int_{{\rm I\!R}^4} {\frac{{{d^4}\omega }}{{{{\left( {2\pi } \right)}^4}}}\int_0^L {dz{H_{\vec \omega }}\left( z \right)\int_0^L {dz'} H_{\vec \omega '}^*\left( {z'} \right)} },
\end{equation}
\begin{equation} \label{eq: X_3 def.}
\mathcal{X}_3=\int_{{\rm I\!R}^4} {\frac{{{d^4}\omega }}{{{{\left( {2\pi } \right)}^4}}}\int_0^L {dz{H_{\vec \omega }}\left( z \right)\int_0^L {dz'} H_{\vec \omega ''}^*\left( {z'} \right)} },
\end{equation}
\begin{equation} \label{eq: X_4 def.}
\mathcal{X}_4=\int_{{\rm I\!R}^3} {\frac{{{d^3}\omega }}{{{{\left( {2\pi } \right)}^3}}}\int_0^L {dzH_{\vec \omega ''}^*\left( z \right)} },
\end{equation}
\begin{equation} \label{eq: X_5 def.}
\mathcal{X}_5=T\int_{{\rm I\!R}^4} {\frac{{{d^5}\omega }}{{{{\left( {2\pi } \right)}^5}}}\int_0^L {dz{H_{\vec \omega }}\left( z \right)\int_0^L {dz'} H_{\vec \omega '''}^*\left( {z'} \right)} },
\end{equation}
\begin{equation} \label{eq: X_1,s def.}
\mathcal{X}_{1,s}=\frac{1}{T}\int {\frac{{{d^3}\omega }}{{{{\left( {2\pi } \right)}^3}}}\int_0^L {dz{H_{\vec \omega ,s}}\left( z \right)} } \int_0^L {dz'} H_{\vec \omega ,s}^*\left( {z'} \right),
\end{equation}
\begin{equation} \label{eq: X_3,s def.}
\mathcal{X}_{3,s} =\int {\frac{{{d^4}\omega }}{{{{\left( {2\pi } \right)}^4}}}\int_0^L {dz{H_{\vec \omega ,s}}\left( z \right)\int_0^L {dz'} H_{\vec \omega '',s}^*\left( {z'} \right)} },
\end{equation} 
\begin{equation} \label{eq: X_1,s,s' def.} 
\mathcal{X}_{1,s,s'}=\frac{1}{T}\int {\frac{{{d^3}\omega }}{{{{\left( {2\pi } \right)}^3}}}\int_0^L {dz{H_{\vec \omega ,s,s'}}\left( z \right)} } \int_0^L {dz'} H_{\vec \omega ,s,s'}^*\left( {z'} \right).
\end{equation}
The following integral representations are found for the $\chi$-coefficients appearing in in the variance of NSNI distortions, (\ref{eq: c_k variance in terms of Chi's}),
\begin{equation} \label{eq: chi_0 final result} 
{\chi _0} = 2\int_0^L {dzf(z)}\int_0^z{dz'}f(z')\xi(z';0), 
\end{equation}
\begin{multline} \label{eq: chi_1 final result}
{\chi _1} =  \frac{2}{{T}}{\mathop{\rm Re}\nolimits} \left\{ \int_{{\rm I\!R}^3}\frac{d^3\omega}{{(2\pi)}^3} \times\right.  \\
\left. {\int_0^L {dz{H_{\vec \omega }}\left( z \right)} \int_0^z {dz'H_{\vec \omega }^*\left( {z'} \right)}\xi(z';0) } \right\},
\end{multline}
\begin{multline} \label{eq: chi_2 final result}
{\chi _2} = 2{\mathop{\rm Re}\nolimits} \left\{ \int_{{\rm I\!R}^4}\frac{d^4\omega}{{(2\pi)}^4} \times\right. \\
\left. {\int_0^L {dz{H_{\vec \omega }}\left( z \right)} \int_0^z {dz'H_{\vec \omega '}^*\left( {z'} \right)} \xi(z';0) } \right\},
\end{multline}
\begin{multline} \label{eq: chi_3 final result}
{\chi _3} = 2{\mathop{\rm Re}\nolimits} \left\{ \int_{{\rm I\!R}^4}\frac{d^4\omega}{{(2\pi)}^4} \times\right. \\
\left. {\int_0^L {dz{H_{\vec \omega }}\left( z \right)} \int_0^z {dz'H_{\vec \omega''}^*\left( {z'} \right)}\xi(z';0)  } \right\},
\end{multline}
\begin{multline} \label{eq: chi_1,s final result}
{\chi _{1,s}} = {\frac{2}{{{T}}}}{\mathop{\rm Re}\nolimits} \left\{ \int_{{\rm I\!R}^3}\frac{d^3\omega}{{(2\pi)}^3} \times\right. \\
\left. {\int_0^L {dz{H_{\vec \omega ,s}}\left( z \right)} \int_0^z {dz'H_{\vec \omega ,s}^*\left( {z'} \right)}\xi(z';0) } \right\},
\end{multline}
\begin{multline} \label{eq: chi'_1,s final result}
{{\chi '}_{1,s}} = {\frac{2}{{{T}}}}{\mathop{\rm Re}\nolimits} \left\{ \int_{{\rm I\!R}^3}\frac{d^3\omega}{{(2\pi)}^3} \times\right. \\
\left. {\int_0^L {dz{H_{\vec \omega ,s}}\left( z \right)} \int_0^z {dz'H_{\vec \omega ,s}^*\left( {z'} \right)}\xi(z';\Omega_s) } \right\},
\end{multline}
\begin{multline} \label{eq: chi_3,s final result} 
{\chi _{3,s}} = 2{\mathop{\rm Re}\nolimits} \left\{ \int_{{\rm I\!R}^4}\frac{d^4\omega}{{(2\pi)}^4} \times\right. \\
\left. {\int_0^L {dz{H_{\vec \omega,s }}\left( z \right)} \int_0^z {dz'H_{\vec \omega'',s}^*\left( {z'} \right)}\xi(z';0)  } \right\},
\end{multline}
where, we have used the shorthand notations
\begin{equation} \label{eq: omega' vect. def.}
\vec \omega ' = \left( {{\omega _4},{\omega _2},{\omega _1} + {\omega _3} - {\omega _4}} \right),
\end{equation}
and
\begin{equation} \label{eq: omega'' vect. def.}
\vec \omega'' = \left( {{\omega _1},{\omega _4},{\omega _3} - {\omega _2}+{\omega _4}} \right).
\end{equation}
Note that $\chi_{0,s}=\chi_0$. The coefficients $\chi_0$ and $\chi_{0,s}$ model the average phase rotation due to nonlinear-signal-noise interactions and will not contribute to the variance of the nonlinear distortions.   
\subsection{Efficient integration}\label{app. subsec. C}
The integral representations derived in the previous subsection for $\mathcal{X}$-coefficients, Eqns. (\ref{eq: X_1 def.})-(\ref{eq: X_1,s,s' def.}), and for $\chi$-coefficients, Eqns. (\ref{eq: chi_0 final result})-(\ref{eq: chi_3,s final result}), have to be evaluated numerically; however, numerical integration of those equations can be still time consuming. In this subsection, we assume EDFA-only optical amplification, and carry out the $z$-integrals. The remaining multidimensional $\omega$ integrals can be efficiently computed by standard Monte Carlo integration. We present the detailed computation only for $\chi_2$ and $\mathcal{X}_2$ to save space. The same procedure can be applied to all other $\mathcal{X}$ and $\chi$-coefficients.  

Let's consider the integral representation of $\chi_2$ as per (\ref {eq: chi_2 final result}), which is rewritten in a slightly different form as 
\begin{equation} \label{eq: chi_2 rewritten with R}
\chi_2 = 2\mathop{\rm Re}\nolimits \{ \int_{{\rm I\!R}^4}\frac{d^4\omega}{{(2\pi)}^4} {R_{\vec \omega, \vec \omega'}}  \},
\end{equation}
where, the integrand function $R_{\vec \omega ,\vec \omega '}$ is
\begin{equation} \label{eq: R_omega def.}
{R_{\vec \omega ,\vec \omega '}} = \int_0^L {dz} {H_{\vec \omega }}\left( z \right){h_{\vec \omega '}}\left( z \right),
\end{equation}
The function $h_{\vec\omega'}(z)$ in (\ref{eq: R_omega def.}) is
\begin{equation} \label{eq: h(z) definition}
{h_{\vec \omega '}}\left( z \right) = \int_0^z {dz'} H_{\vec \omega '}^*\left( {z'} \right)\xi \left( {z';0} \right).
\end{equation}
We also define the set of auxiliary variables $\{\psi_n\}$, for $n=1,\cdots, N_s$ as
\begin{equation} \label{eq: psi definition}
{\psi _n} = \sum\limits_{j = 0}^{n - 1} {{\xi_j(0)}}.
\end{equation}
Now we substitute (\ref{eq: xi(z,omega) simplified formula}) into (\ref{eq: h(z) definition}), and use (\ref{eq: psi definition}). After doing some algebra we obtain
\begin{multline} \label{eq: h(z) as h_m + integral}
{h_{\vec \omega '}}\left( z \right) = {h_{\vec \omega ',m}} + {\psi _{m}}\int_{{z_{m - 1}}}^z {dz'} H_{\vec \omega '}^*\left( {z'} \right) \\
{z_{m - 1}} \leq z < {z_m},m = 1, \ldots ,{N_s},
\end{multline}
where,
\begin{equation} \label{eq: h_omega',m}
{h_{\vec \omega ',m}} = \sum\limits_{j = 0}^{m - 1} {{\psi _j}r_{\vec \omega ',j}^*},
\end{equation}
and
\begin{equation} \label{eq: r_omega',m}
{r_{\vec \omega,m}} = \int_{{z_{m - 1}}}^{{z_m}} {dz'{H_{\vec \omega}}\left( {z'} \right)}.
\end{equation}
If we assume that $\beta_2(z) = {\beta _{2,m}}$, and $\alpha(z) = {\alpha_m}$ for $z_{m - 1}\leq z < {z_m}$, The integration in (\ref{eq: r_omega',m}) can be carried out to obtain
\begin{equation} \label{eq: r closed form}
r_{\vec \omega ,m} = e^{-\delta_{m-1}}\frac{{{e^{i{\varphi _{\vec \omega ,m}}{z_{m - 1}}}} - {\eta _m}{e^{i{\varphi _{\vec \omega ,m}}{z_m}}}}}{{{\alpha _m} - i{\varphi _{\vec \omega ,m}}}}{\Pi _{\vec \omega }},
\end{equation}
where, the phase factor due to dispersion is
\begin{equation} \label{eq:}
\varphi _{\vec \omega ,m} = {\beta _{2,m}}\left( {{\omega _2} - {\omega _1}} \right)\left( {{\omega _2} - {\omega _3}} \right).
\end{equation}
The field span loss is 
\begin{equation} \label{eq: eta_m def.}
\eta_m = {e^{ - {\alpha _m}({{z_m} - {z_{m - 1}}})}}.
\end{equation}
Now we rewrite (\ref{eq: R_omega def.}) as a sum over spans as follows\footnote{Note that $L = z_{N_s}$.}
\begin{equation} \label{eq: R_omega span-by-span}
R_{\vec \omega ,\vec \omega '} = \sum\limits_{m = 1}^{{N_s}} {\int_{{z_{m - 1}}}^{{z_m}} {dz} {H_{\vec \omega }}\left( z \right){h_{\vec \omega '}}\left( z \right)},
\end{equation}
and substitute (\ref{eq: h(z) as h_m + integral}) into (\ref{eq: R_omega span-by-span}). The following formula is derived for $R_{\vec \omega ,\vec \omega'}$
\begin{equation} \label{eq: R_omega fast formulation}
{R_{\vec \omega ,\vec \omega '}} = \sum\limits_{m = 1}^{{N_s}} {{r_{\vec \omega ,m}}} {h_{\vec \omega ',m}} + \sum\limits_{m = 1}^{{N_s}} {{\psi _m}} {I_{\vec \omega ,\vec \omega ',m}},
\end{equation}
where,
\begin{equation} \label{eq: I_omega_omega',m def.}
I_{\vec \omega ,\vec \omega', m} = \int_{{z_{m - 1}}}^{{z_m}} {dz} {H_{\vec \omega }}\left( z \right)\int_{{z_{m - 1}}}^z {dz'} H_{\vec \omega '}^*(z').
\end{equation}
The integral in the right hand side of (\ref {eq: I_omega_omega',m def.}) is calculated in a straightforward manner, to obtain 
\begin{multline} \label{eq: I_m explicit computation}
{I_{\vec \omega ,\vec \omega ',m}} = {\Pi _{\vec \omega }}{\Pi _{\vec \omega '}}\frac{{{e^{ - 2{\delta _{m - 1}} + i\left( {{\varphi _{\vec \omega ,m}} - {\varphi _{\vec \omega ',m}}} \right){z_{m - 1}}}}}}{{{\alpha _m} + i{\varphi _{\vec \omega ',m}}}} \times \\
\{ \frac{{1 - {\eta _m}{e^{i{\varphi _{\vec \omega ,m}}\left( {{z_m} - {z_{m - 1}}} \right)}}}}{{{\alpha _m} - i{\varphi _{\vec \omega ,m}}}} - \\
\frac{{1 - \eta _m^2{e^{i(\varphi_{\vec\omega,m}-\varphi_{\vec\omega',m})(z_m-z_{m-1})}}}}{{2{\alpha _m} - ({\varphi _{\vec \omega ,m}} - \varphi_{\vec\omega',m})}} \}.
\end{multline}
Putting it all together, the coefficient $\chi_2$ can be evaluated by the following four-dimensional integral 
\begin{multline} \label{eq: chi_2 efficient integral representation}
\chi_2 = \\
2\operatorname{Re} \left\{ {\int_{V} {\frac{{{d^4}\omega }}{{{{\left( {2\pi } \right)}^4}}}} } \right.\left. {\left[ {\sum\limits_{m = 1}^{{N_s}} {{r_{\vec \omega ,m}}} {h_{\vec \omega',m}} + \sum\limits_{m = 1}^{{N_s}} {{\psi _m}} {I_{\vec \omega ,\vec \omega ',m}}} \right]} \right\},
\end{multline}
where, $V=\left\{ {\left( {{\omega _1},{\omega _2},{\omega _3},{\omega _4}} \right) \in {{\rm I\!R}^4}\left| {{\Pi _{\vec \omega }}{\Pi _{\vec \omega '}} \ne 0} \right.} \right\}$ is a four-dimensional subset of the four-dimensional real space\footnote{Note that for Nyquist pulses assumed in this work, (cf. (\ref{eq: sinc spectrum})),  the function $\Pi_{\vec\omega}$ takes on values in the set $\{0,T^2\}$, thus acting like a gating function.}. The integral in (\ref {eq: chi_2 efficient integral representation}) can be efficiently computed by Monte Carlo sampling of the four-dimensional region $V$. 

A similar procedure can be applied to efficiently compute all other $\chi$-coefficients. In order to compute $\chi_3$ and $\chi_{3,s}$, we have to replace $\vec\omega'$ with $\vec\omega''$ in (\ref {eq: chi_2 efficient integral representation}). To compute $\chi_{3,s}$, we also have to replace $H_{\vec\omega}$ in (\ref{eq: r_omega',m}) with $H_{\vec\omega,s}$. In order to compute $\chi_1$, $\chi_{1,s}$ and $\chi'_{1,s}$ we have to replace $\vec\omega'$ with $\vec\omega$ in (\ref{eq: R_omega def.})-(\ref {eq: chi_2 efficient integral representation}). In this case the right hnd side of (\ref{eq: I_m explicit computation}) is considerbaly simplified, and we obtain 
\begin{equation} \label{eq: I_m explicit computation, omega'=omega}
\operatorname{Re} \left\{ {{I_{\vec \omega ,\vec \omega ,m}}} \right\} = \frac{1}{2}{\left| {{r_{\vec \omega ,m}}} \right|^2}.
\end{equation}
For computing $\chi'_{1,s}$ we also have to replace $\xi_j(0)$ with $\xi_j(\Omega_s)$ in (\ref{eq: psi definition}).

The same approach can be applied, with much less pain, to efficiently integrate $\mathcal{X}$-coefficients. For instance, the following expression is derived for $\mathcal{X}_2$ 
\begin{equation} \label{eq: X_2 efficient integration}
\mathcal{X}_2=\int_{{\rm I\!R}^4} {\frac{{{d^4}\omega }}{{{{\left( {2\pi } \right)}^4}}}} \left( {\sum\limits_{m = 1}^{{N_s}} {{r_{\vec \omega ,m}}} } \right)\left( {\sum\limits_{m = 1}^{{N_s}} {r_{\vec \omega ',m}^*} } \right).
\end{equation}

%\end{IEEEbiography}


\end{document}